\documentclass[preprint]{aastex63}
\usepackage{xspace}
\usepackage{graphicx}
\usepackage{natbib}
\usepackage{url}
\usepackage{bm}
\usepackage{amsmath}

\newcommand{\degree}{\ensuremath{\,^\circ}}

\newcommand{\ghz}{\ensuremath{\,{\rm GHz}}}
\newcommand{\kel}{\ensuremath{\,{\rm K}}}

\newcommand{\percc}{\ensuremath{\,{\rm cm^{-3}}}}

\newcommand{\kpc}{\ensuremath{\,{\rm kpc}}}
\newcommand{\pc}{\ensuremath{\,{\rm pc}}}
\newcommand{\kms}{\ensuremath{\,{\rm km\, s^{-1}}}}

\newcommand{\minute}{\,minute}

\newcommand{\jyb}{\ensuremath{\rm \,Jy\,beam^{-1}}}

\newcommand{\jya}{\ensuremath{\rm \,Jy\,arcsec^{-2}}}
\newcommand{\mjya}{\ensuremath{\rm \,mJy\,arcsec^{-2}}}

\newcommand{\kmspc}{\ensuremath{\,{\rm km\, s^{-1}\, pc^{-1}}}}

\newcommand{\msarcsec}{\ensuremath{\,{\rm m\, s^{-1}\, arcsec^{-1}}}}

\newcommand{\hna}{\ensuremath{\langle {\rm Hn}\alpha \rangle}}

\newcommand{\expo}[1]{\ensuremath{10^{#1}}}

\newcommand{\hi}{H\,{\sc i}}

\newcommand{\hii}{H\,{\sc ii}}

\newcommand{\nii}{N\,{\sc ii}}
\newcommand{\neii}{Ne\,{\sc ii}}

\newcommand{\ngc}[1]{NGC\thinspace #1}

\newcommand{\vect}[1]{\bm{#1}}
\newcommand\urltilda{\kern -.15em\lower .7ex\hbox{\~{}}\kern .04em}
\newcommand{\lsim}{\ensuremath{\lesssim}}

\begin{document}

\title{Discovery of a New Population of Galactic \hii\ Regions with
  Ionized Gas Velocity Gradients}

\author[0000-0002-2465-7803]{Dana S. Balser}
\affiliation{National Radio Astronomy Observatory, 520 Edgemont Rd.,
  Charlottesville, VA 22903, USA}

\author[0000-0003-0640-7787]{Trey V. Wenger}
\affiliation{Dominion Radio Astrophysical Observatory, Herzberg
  Astronomy and Astrophysics Research Centre, National Research
  Council, P.O.  Box 248, Penticton, BC V2A 6J9, Canada}

\author[0000-0001-8800-1793]{L. D. Anderson} \affiliation{Department
  of Physics and Astronomy, West Virginia University, Morgantown, WV
  26505, USA} \affiliation{Center for Gravitational Waves and
  Cosmology, West Virginia University, Morgantown, Chestnut Ridge
  Research Building, Morgantown, WV 26505, USA} \affiliation{Adjunct
  Astronomer at the Green Bank Observatory, P.O. Box 2, Green Bank, WV
  24944, USA}

\author[0000-0002-7045-9277]{W. P. Armentrout} \affiliation{Green Bank
  Observatory, P.O. Box 2, Green Bank, WV 24944, USA}

\author[0000-0003-4866-460X]{T. M. Bania} \affiliation{Institute
  for Astrophysical Research, Astronomy Department, Boston University,
  725 Commonwealth Avenue, Boston, MA 02215, USA}

\author[0000-0003-0235-3347]{J. R. Dawson} \affiliation{Department of
  Physics and Astronomy and MQ Research Centre in Astronomy,
  Astrophysics, and Astrophotonics, Macquarie University, NSW 2109,
  Australia} \affiliation{Australia Telescope National Facility, CSIRO
  Astronomy and Space Science, P.O. Box 76, Epping, NSW 1710,
  Australia}

\author[0000-0002-6300-7459]{John M. Dickey} \affiliation{School of
  Natural Sciences, University of Tasmania, Hobart, TAS 7001,
  Australia}

\begin{abstract}

  We investigate the kinematic properties of Galactic \hii\ regions
  using radio recombination line (RRL) emission detected by the
  Australia Telescope Compact Array (ATCA) at 4--10\ghz\ and the
  Jansky Very Large Array (VLA) at 8--10\ghz.  Our \hii\ region sample
  consists of 425 independent observations of 374 nebulae that are
  relatively well isolated from other, potentially confusing sources
  and have a single RRL component with a high signal-to-noise ratio.
  We perform Gaussian fits to the RRL emission in
  position-position-velocity data cubes and discover velocity
  gradients in 178 (42\%) of the nebulae with magnitudes between 5 and
  200\msarcsec. About 15\% of the sources also have a RRL width
  spatial distribution that peaks toward the center of the nebula.
  The velocity gradient position angles appear to be random on the sky
  with no favored orientation with respect to the Galactic Plane.  We
  craft \hii\ region simulations that include bipolar outflows or
  solid body rotational motions to explain the observed velocity
  gradients.  The simulations favor solid body rotation since, unlike
  the bipolar outflow kinematic models, they are able to produce both
  the large, $> 40$\msarcsec, velocity gradients and also the RRL
  width structure that we observe in some sources.  The bipolar
  outflow model, however, cannot be ruled out as a possible
  explanation for the observed velocity gradients for many sources in
  our sample.  We nevertheless suggest that most \hii\ region
  complexes are rotating and may have inherited angular momentum from
  their parent molecular clouds.
  
\end{abstract}


\section{Introduction}\label{sec:intro}

\hii\ regions are the zones of ionized gas surrounding young, OB-type
stars. Feedback from these massive stars is an important mechanism in
galaxy formation and evolution \citep[e.g.,][]{veilleux05}.  For
example, high-mass stars can produce radiation-driven outflows that
both inhibit and stimulate star formation in their vicinities
\citep[e.g.,][]{ali18}.  Furthermore, the internal motions of \hii\
regions constrain models of star formation and evolution
\citep[e.g.,][]{dalgleish18}.  For example, high frequency RRLs that
probe the high density gas within the ultra-compact \hii\ region
G10.6$-$0.4 support evidence of an ionized accretion flow
\citep{keto06}.  RRL observations toward K3-50 reveal a velocity
gradient across the nebula.  This together with a bipolar morphology
from the radio continuum emission indicates that this source is
undergoing a high-velocity bipolar outflow \citep{depree94, balser01}.
RRL position-velocity diagrams of \ngc{6334A} reveal both a bipolar
outflow and rotation in the ionized gas \citep{depree95, balser01}.
There is evidence that high-mass stars form in binaries or multiple
star systems more often than low-mass stars \citep{mermilliod01,
  okamoto03}, creating massive accretion flows that overcome radiation
and thermal pressure \citep{keto06}.  Therefore \hii\ region bipolar
outflows may arise from multiple OB-type stars.  Most \hii\ region
studies of this type, however, focus on one or a few sources.  To our
knowledge there have not been any surveys of the kinematic motions
within \hii\ regions for a large number of sources.

\hii\ regions occupy intermediate spatial scales between giant
molecular cloud (GMC) star forming complexes and individual
protostars, both of which have a rich literature of kinematics
studies.  For example, on the large---tens of parsecs---scale both
\hi\ and CO tracers have been used to explore the angular momentum in
the largest star forming structures \citep[e.g.,][]{fleck81,
  arquilla85, imara11}. The kinematic structure of molecular cores and
protostars which probe the small---$\lsim 0.05$\pc---scale have been
investigated using dense molecular gas tracers
\citep[e.g.,][]{goodman93, caselli02, chen07, tobin11}.  Young,
ultra-compact \hii\ regions have similar spatial scales with sizes
less than 0.1\pc\ \citep[e.g.,][]{churchwell02}.  But over time the
warm, \expo{4}\kel, plasma causes the \hii\ region to expand creating
a ``classical'' \hii\ region that can span several parsecs, filling in
the gap between GMCs and protostars.  Photo-dissociation regions can
also be exploited to probe the kinematics on these spatial scales
\citep[e.g.,][]{luisi21}.

The lack of \hii\ region kinematic surveys is in part due to
limitations of the available instrumentation of past facilities.
Optical observations of H$\alpha$ typically have the necessary
sensitivity and spatial resolution but often lack the spectral
resolution needed for detailed kinematic studies.  There are some
exceptions \citep[e.g., see][]{russeil16}.  There are some
extragalactic studies of \hii\ region kinematics, but they typically
do not have the angular resolution to probe individual star formation
regions \citep[e.g., see][]{bordalo09, torres-flores13, bresolin20}.
Since H$\alpha$ emission is attenuated by dust the entire volume of
ionized gas is not always sampled, therefore limiting the usefulness
of this tracer.  Fine structure lines at infrared wavelengths can
penetrate the dust.  For example, \nii\ 122\micron\
\citep{lebouteiller12} and \neii\ 12.8\micron\ \citep{jaffe05, zhu06}
trace the dense ionized gas found in \hii\ regions.  To our knowledge,
however, there have not been any infrared studies of the kinematics
for a large sample of \hii\ regions.  In contrast, RRLs are optically
thin tracers with sufficient spectral resolution, but they are often
observed using single-dish telescopes with poor angular resolution.
Radio interferometers can significantly improve the angular resolution
but they are typically not sensitive to the more diffuse emission that
exists in many \hii\ regions.  Here we use RRL data from two new
interferometer surveys that mitigate some of these weaknesses to
investigate \hii\ region kinematics from a large sample of nebulae.

\section{The Data}\label{sec:data}

Data are taken from two recent \hii\ region RRL surveys: the Southern
\hii\ Region Discovery Survey (SHRDS) using the ATCA \citep{wenger21},
and the \hii\ region VLA RRL survey \citep{wenger19a}.  As the first
\hii\ region RRL surveys using interferometer observations toward a
large number of sources, they have sufficient spatial resolution to at
least partially resolve many Galactic \hii\ regions.  Moreover, the
flexibility of their spectrometers allows many RRLs to be
simultaneously observed and averaged together to increase the
signal-to-noise ratio \citep[e.g.,][]{balser06}.  Therefore, for the
first time we have \hii\ region RRL surveys that have both the
sensitivity and spatial resolution at radio wavelengths to measure the
internal \hii\ region kinematics in a large sample.
Table~\ref{tab:stats} summarizes the statistics for these surveys.
Because each interferometer field may contain multiple \hii\ regions,
we detect RRL and continuum emission from more \hii\ regions than are
targeted and also have some duplicate observations.

\begin{deluxetable}{lrrrl}
\tabletypesize{\scriptsize}
\tablecaption{\hii\ Region Survey Statistics \label{tab:stats}}
\tablehead{
\colhead{Statistic} & \colhead{ATCA} & \colhead{VLA} &  \colhead{Total} &
\colhead{Comment} 
}
\startdata
$N_{\rm tot}$ & 965 (669) & 118 (117) & 1083 (786) & Total number of sources in the survey. \\
$N_{\rm kin}$   & 377 (327) &  48 (47) &  425 (374) & Number of sources suitable for kinematic analysis. \\
$N_{\rm grad}$ & 166 (150) &  12 (12) & 178 (162) & Number of sources with a velocity gradient. \\
$f_{\rm grad}$ & 44 (46) & 25 (26) & 42 (43) & Percentage of suitable sources with a velocity gradient: $N_{\rm grad}/N_{\rm kin}$. \\
\enddata
\tablecomments{Counts correspond to independent detections toward
  the ATCA and VLA images.  Since each image may contain multiple
  \hii\ regions there are some duplicate observations.  In parentheses
  we list the unique source counts.}
\end{deluxetable}

\subsection{Australia Telescope Compact Array}\label{sec:atca}

The SHRDS is an ATCA 4--10\ghz\ radio continuum and RRL survey to find
\hii\ regions in the Galactic zone
$259\degree < \ell < 344\degree, |b| < 4\degree$ \citep{brown17,
  wenger19b, wenger21}.  The SHRDS is an extension of the HRDS in the
northern sky that used the Green Bank Telescope (GBT) and also the
Arecibo telescope to observe \hii\ region candidates selected from
public infrared and radio continuum data \citep{bania10, anderson11,
  bania12, anderson15, anderson18}.  These recent \hii\ region RRL
surveys have doubled the number of known \hii\ regions in the Milky
Way.  The SHRDS full catalog consists of 436 new Galactic \hii\
regions and 206 previously known nebulae \citep{wenger21}.  Here we
consider every continuum detection in the SHRDS images for our initial
analysis.  This consists of 965 independent continuum measurements.

Since the ATCA is a compact array, the short baselines are better able
to recover emission from the extended parts of an \hii\ region.  The
SHRDS employed the most compact antenna configuration, H75, together
with the larger H168 configuration to probe a range of spatial scales.
Moreover, the observations were made at different hour angles to
provide good coverage in the uv-plane \citep[see Figure 3
in][]{wenger19b}.  The half-power beam-width (HPBW) spatial resolution
is $\sim 90$\arcsec\ and the maximum recoverable scale is
$\sim 265$\arcsec.  The ATCA Compact Array Broadband Backend (CABB)
can tune simultaneously to 20 different RRLs across the 4--10\ghz\
band.  Typically, about 18 RRL transitions between
H88$\alpha$--H112$\alpha$ were suitable to average together (stack) to
produce a final spectrum, \hna.  To do this we first re-grid the
smoothed RRL data cubes to a common spectral and angular resolution,
and then average the smoothed, regridded data cubes.

We select sources from the SHRDS that have the following criteria.
(1) They must be either well isolated or only slightly confused, with
a single continuum emission peak.  All sources with quality factor A
or B meet this criterion \citep[see][]{wenger21}.  (2) Spectra toward
some \hii\ regions contain multiple RRL components which may be due to
more than one distinct \hii\ region.  So we restrict our analysis to
sources with only single RRL components.  (3) We require good
sensitivity to infer kinematic structure across the \hii\ region.
Therefore we only select sources with a peak RRL signal-to-noise
ratio, ${\rm SNR} \ge 10$.  Applying these criteria to the SHRDS
yields 377 independent observations of 327 unique sources for
kinematic analysis.

The SHRDS provides a very coarse measure of the kinematic structure in
\hii\ regions.  Most nebulae are resolved with only a few synthesized
beams across the detected nebula emission.  By design the ATCA
observations, taken in the most compact configurations, do not provide
very detailed images; the synthesized beam is not significantly
smaller than the primary beam (HPBW = 225\arcsec).  This is because
for the SHRDS science goals surface brightness sensitivity is more
important than spatial resolution.  Nevertheless, here we will show
that the sensitivity and spatial resolution are sufficient to roughly
characterize the global properties of the internal kinematics of \hii\
regions.

\subsection{Jansky Very Large Array}\label{sec:vla}

The goal of the \hii\ region VLA 8--10\ghz\ RRL survey was to measure
the metallicity throughout the Galactic disk in the northern sky
\citep{wenger19a}. Accurate electron temperatures were derived using
the RRL to continuum intensity ratio.  Since metals act as coolants
within the ionized gas, the electron temperatures are a proxy for
metallicity \citep{rubin85}.  A total of 82 nebulae were targeted.
Multiple sources were sometimes detected within the primary beam and
therefore here we consider every continuum detection in the VLA
images.  This consists of 118 independent continuum measurements.

Data were taken in the most compact VLA D configuration using
$\sim 10$\minute\ ``snapshot'' integrations.  The uv-plane was
therefore not well sampled and the resulting RRL images are not very
sensitive to any extended \hii\ region emission.  The HPBW spatial
resolution is 15\arcsec\ and the maximum recoverable scale is
145\arcsec.  The Wideband Interferometric Digital ARchitecture (WIDAR)
correlator in the 8-bit sampler mode can tune simultaneously to 8 RRL
transitions between H86$\alpha$--H93$\alpha$ across the 8--10\ghz\
band.  Since the H86$\alpha$ transition is confused with a higher
order RRL only 7 RRLs are stacked to produce the \hna\ spectrum.

We follow the same selection criteria as discussed in
Section~\ref{sec:atca} for our kinematic analysis.  We therefore only
choose relatively isolated nebulae that have a single RRL component
with a ${\rm SNR} \ge 10$.  A total of 48 independent observations of
47 unique sources meet these criteria.

\section{\hii\ Region Kinematics}\label{sec:kinematics}

Random gas motions of the plasma produced by photoionziation within
\hii\ regions should be well described by a Maxwell-Boltzmann
distribution.  Such motions produce Gaussian spectral line profiles
for optically thin gas via the Doppler effect.  There is some evidence
for small deviations from a Maxwell-Boltzmann distribution in \hii\
regions \citep[e.g.,][]{nicholls12}, but this should not significantly
affect our interpretation of \hii\ region kinematics.  Observations of
spectral lines from ionized gas tracers indicate line widths
significantly broader than the thermal width, even on small spatial
scales \citep{ferland01, beckman04}.  These non-thermal line widths
have been attributed to MHD waves \citep[e.g.,][]{mouschovias75} or
turbulence \citep[e.g.,][]{morris74} which also produce Gaussian line
profiles via the Doppler effect.  Bulk motions of the gas produced by
dynamical effects may alter the line shape. For example, an expanding
\hii\ region could produce a double-peaked RRL profile if the nebula
is resolved, or a square-wave RRL profile for an unresolved source
\citep[e.g.,][]{balser97}.  But analysis of \hna\ spectra toward
nebulae in the ATCA and VLA surveys typically reveal a single Gaussian
line profile.  There are six nebulae (3\%) with significant
non-Gaussian line wings or double-peaked profiles that are not
included in our analysis.

\begin{figure}
  \centering
  \includegraphics[angle=0,scale=0.45]{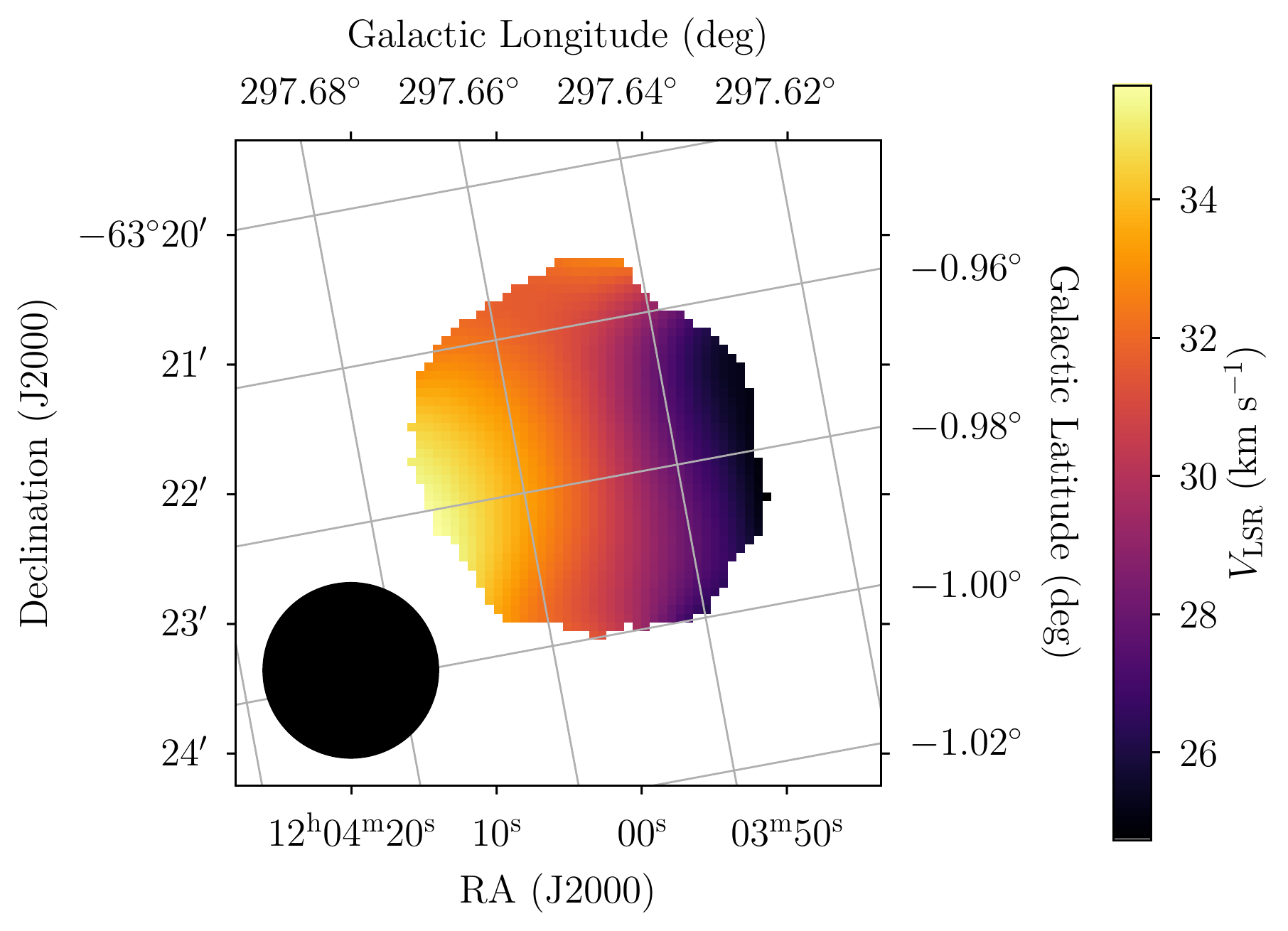}
  \includegraphics[angle=0,scale=0.45]{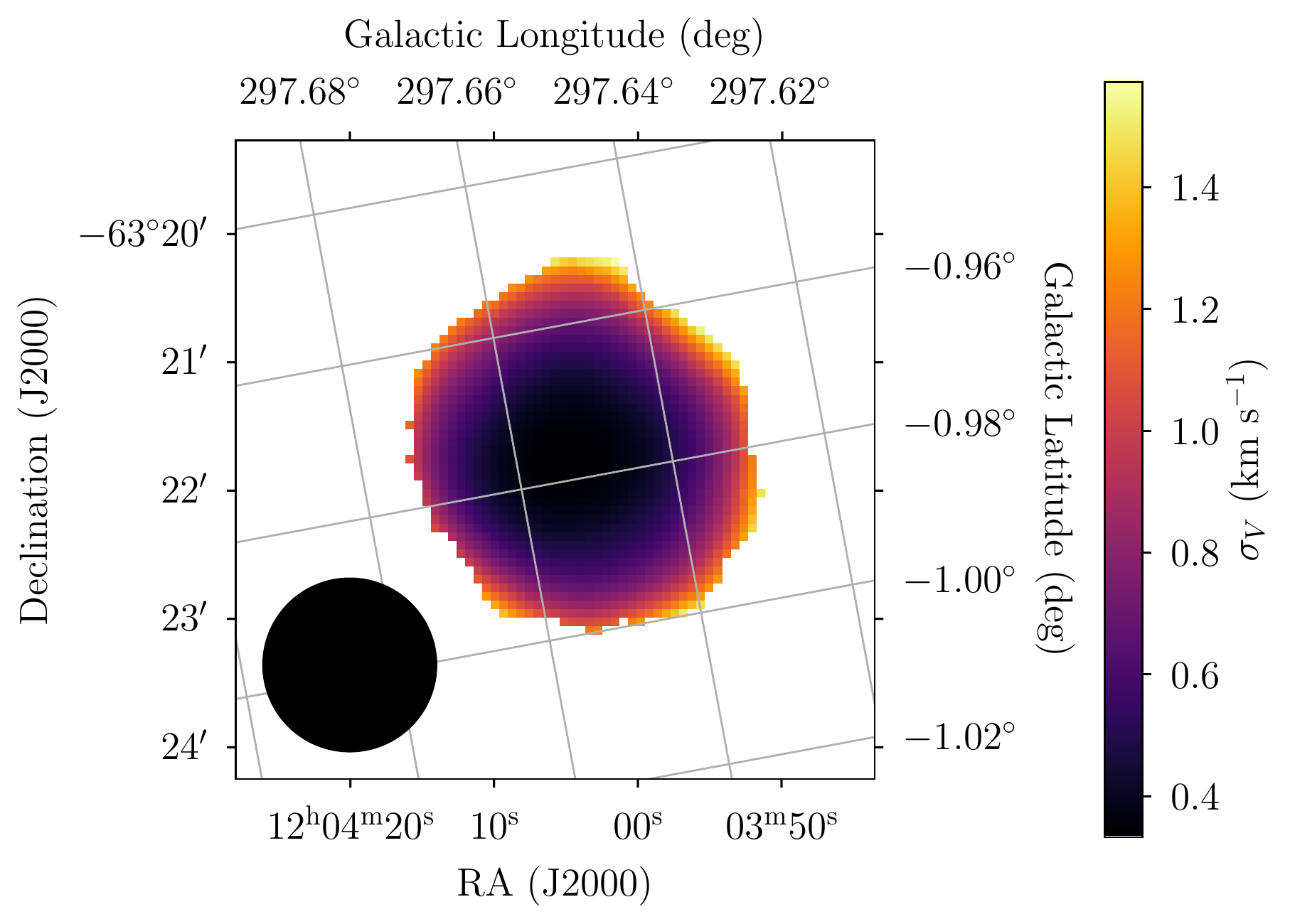}
  \includegraphics[angle=0,scale=0.45]{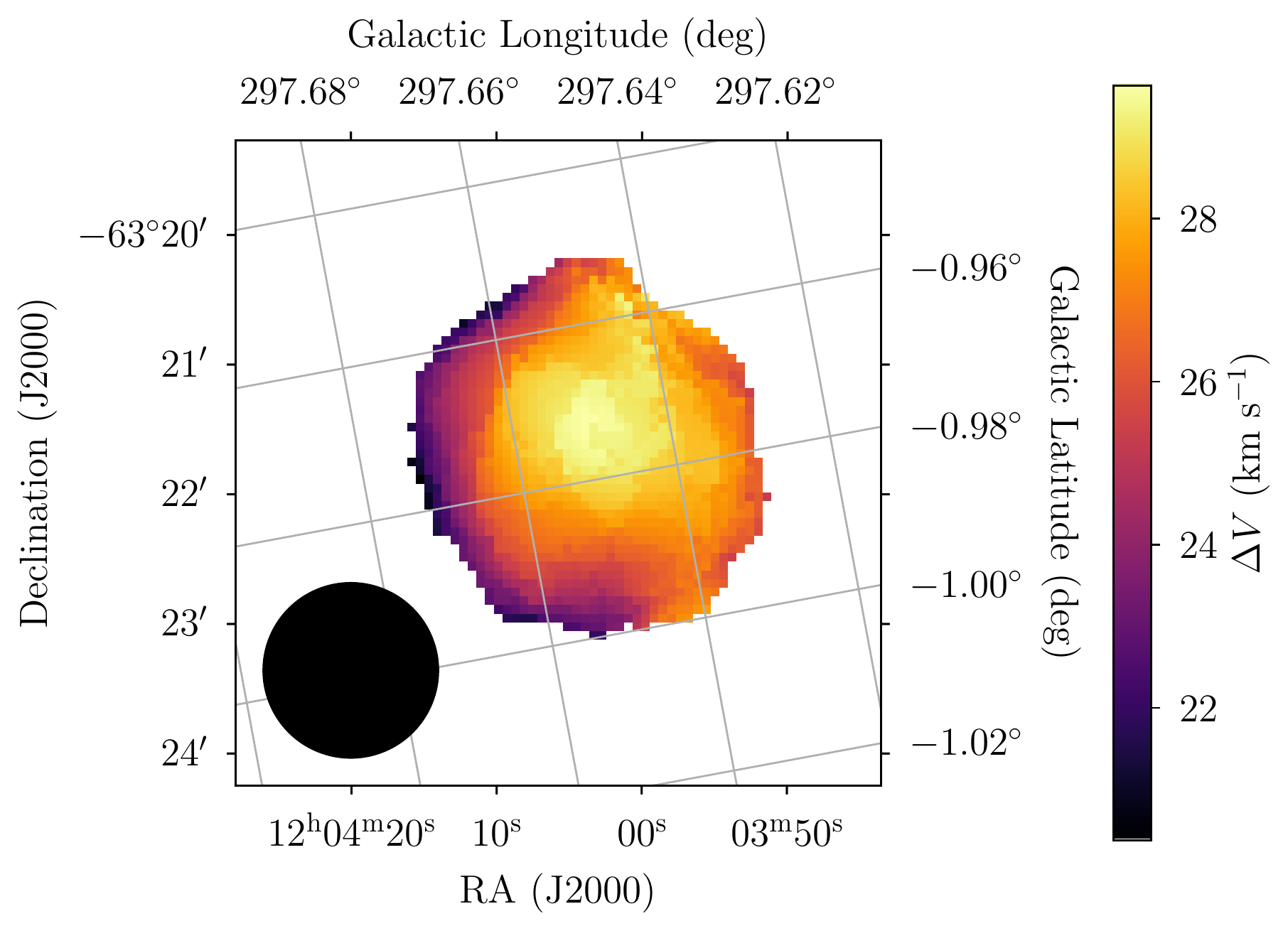}
  \includegraphics[angle=0,scale=0.45]{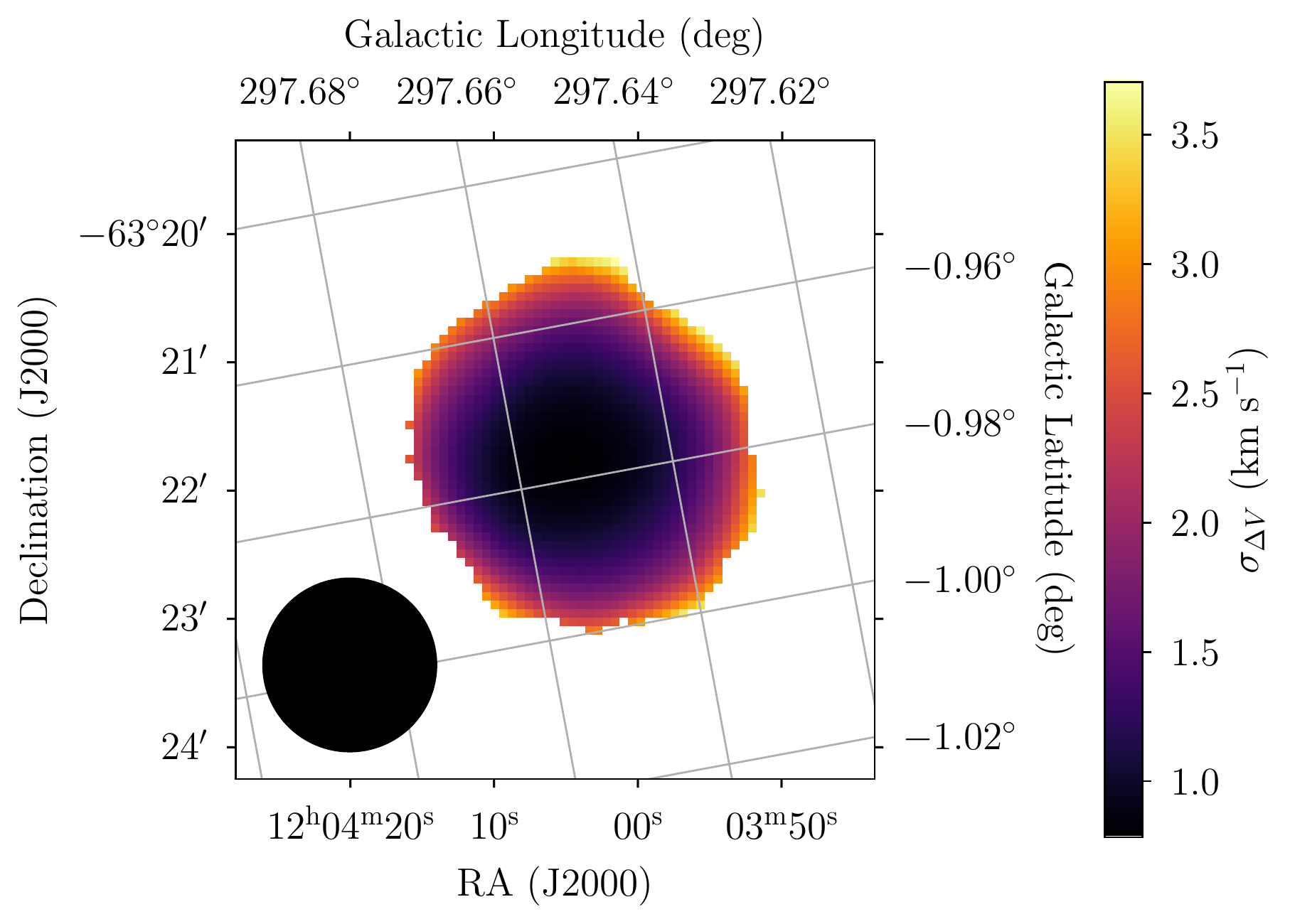}
  \caption{\hna\ spectrum RRL Gaussian center velocity, $V_{\rm LSR}$,
    and line width, $\Delta V$, parameter images of G297.651$-$00.973,
    observed with the ATCA.  {\it Top:} Gaussian fit of the center
    velocity, $V_{\rm LSR}$, and the associated error.  {\it Bottom:}
    Gaussian fit of the FWHM line width, $\Delta V$, and the
    associated error.  The boundaries of the fitted region are defined
    by a watershed segmentation algorithm that identifies all pixels
    with emission associated with an emission peak \citep[for details
    see][]{wenger21}.  Outside the watershed region the images have a
    poor signal-to-noise ratio and there are no Gaussian RRL fits.
    The synthesized HPBW is shown by the black circle in the bottom
    left-hand corner of the image.}
\label{fig:grad-atca}
\end{figure}

\begin{figure}
  \centering
  \includegraphics[angle=0,scale=0.45]{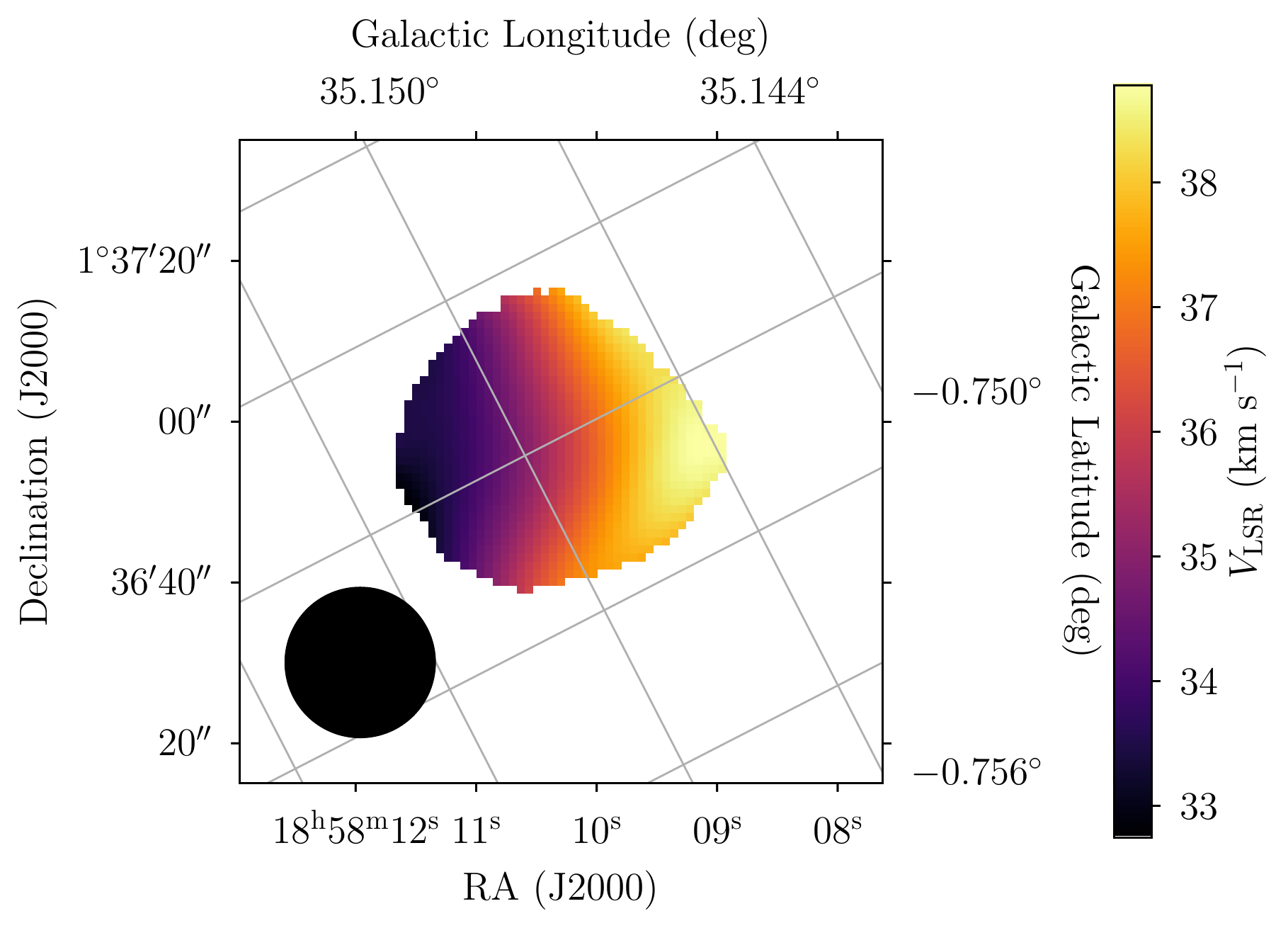}
  \includegraphics[angle=0,scale=0.45]{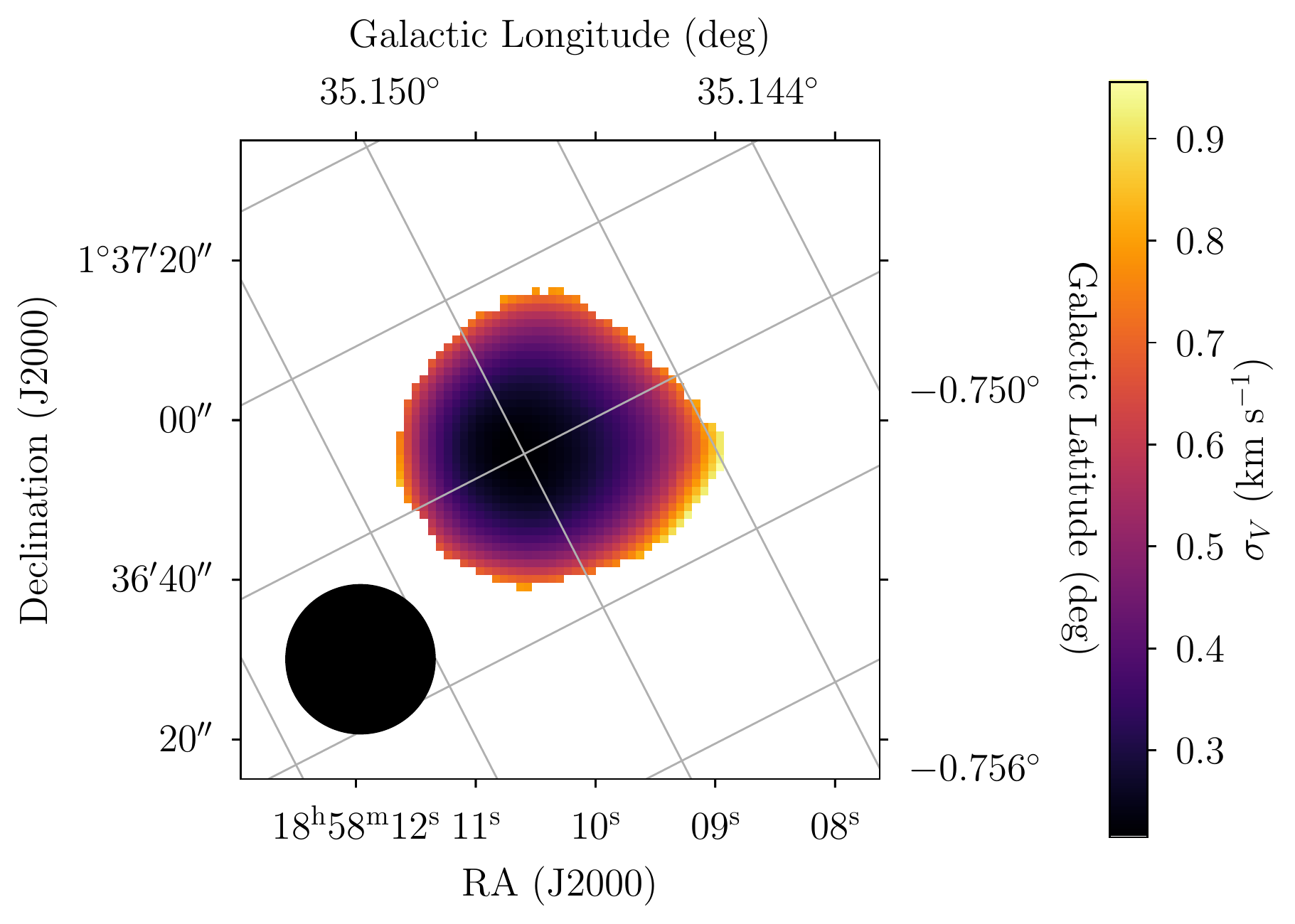}
  \includegraphics[angle=0,scale=0.45]{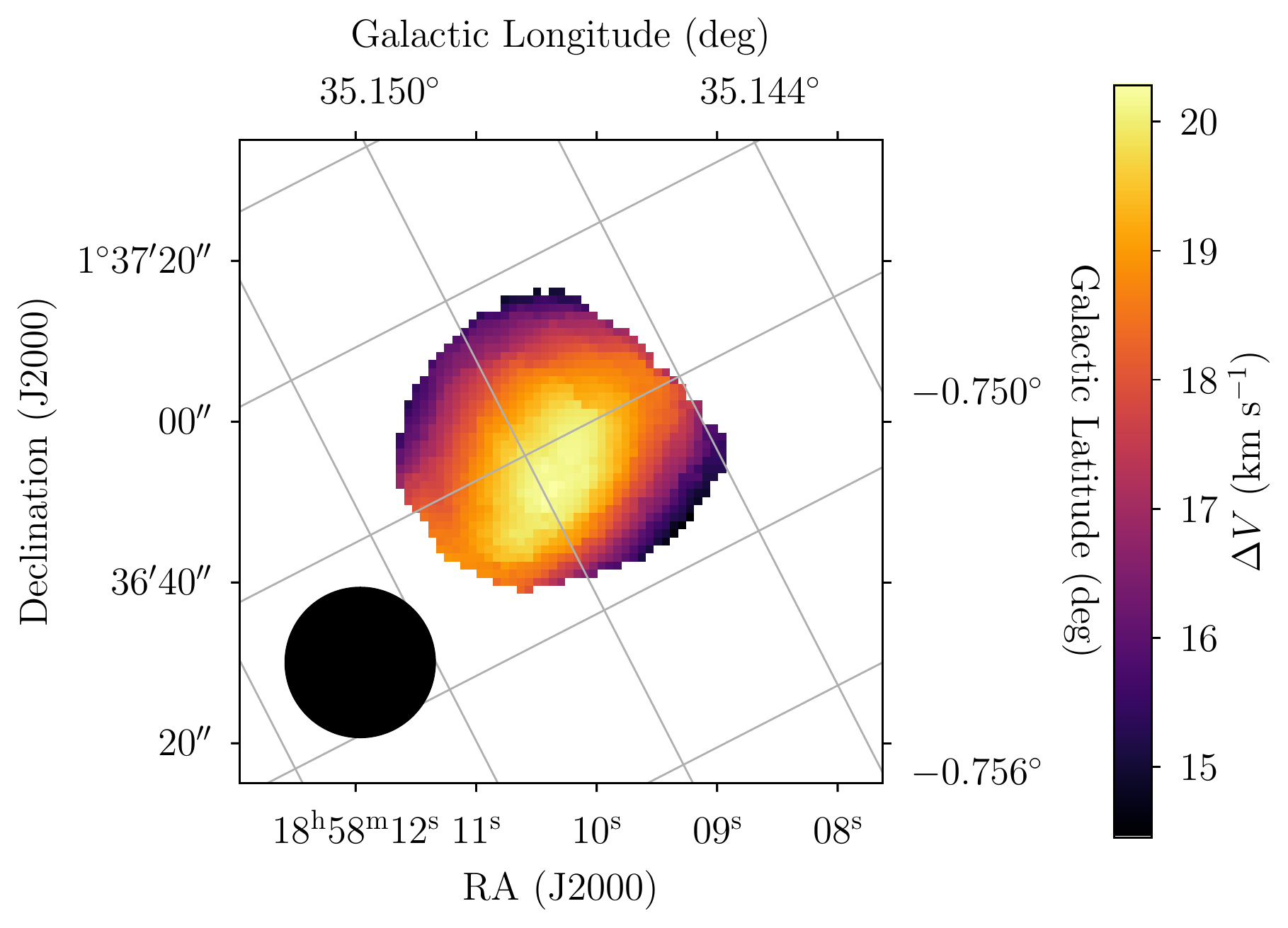}
  \includegraphics[angle=0,scale=0.45]{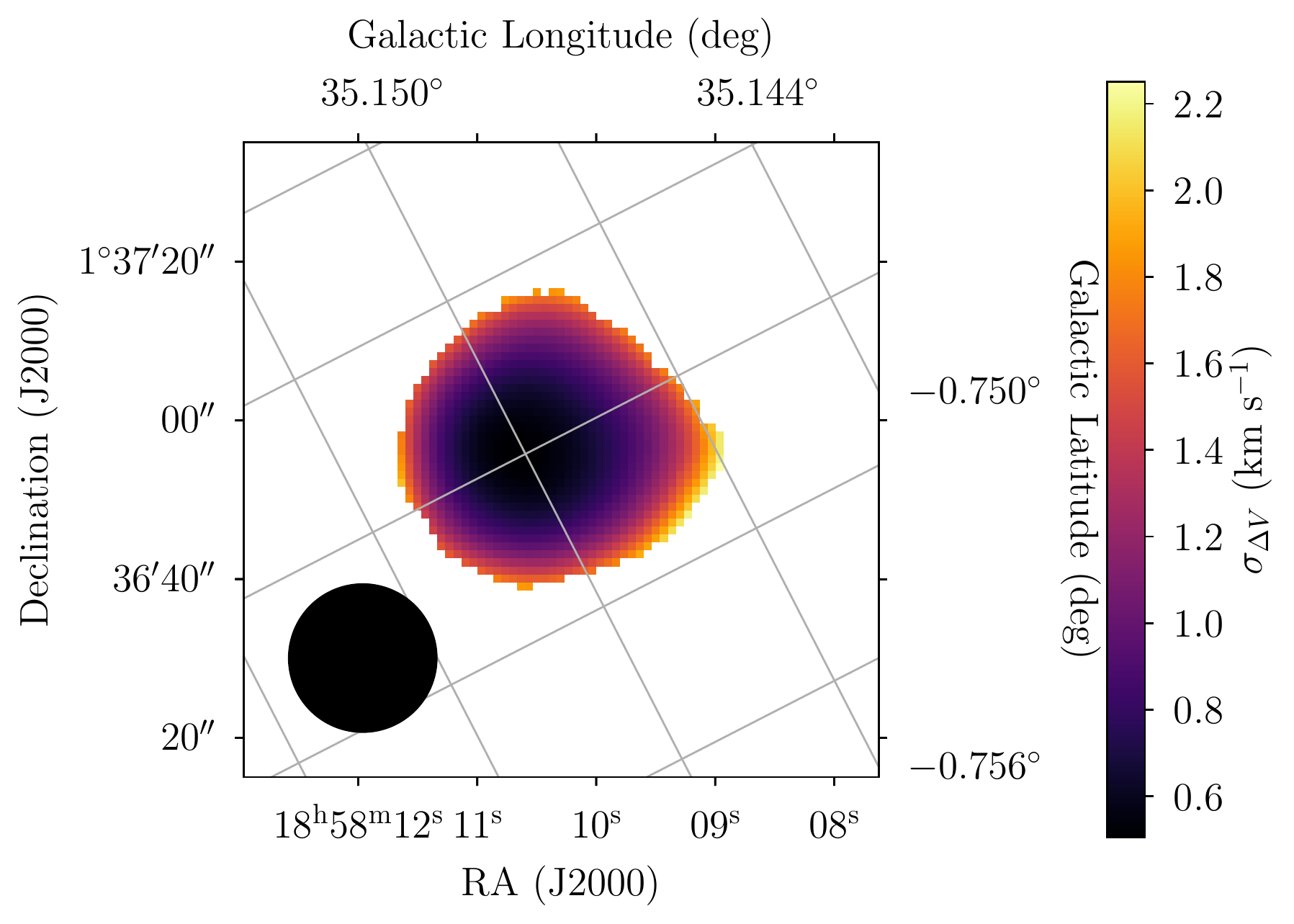}
  \caption{\hna\ spectrum RRL Gaussian center velocity, $V_{\rm LSR}$,
    and line width, $\Delta V$, parameter images of G035.126$-$0.755
    observed with the VLA.  See Figure~\ref{fig:grad-atca} for
    details.}
\label{fig:grad-vla}
\end{figure}

To investigate the \hii\ region kinematics, we therefore perform
single component, Gaussian fits to \hna\ profiles for each spectral
pixel (spaxel) in the RRL data cubes using the Levenberg-Markwardt
\citep{markwardt09} least-squares method.  Figures~\ref{fig:grad-atca}
and \ref{fig:grad-vla} show the results of these fits for
G297.651$-$00.973 and G035.126$-$0.755 taken from the ATCA and VLA
surveys, respectively.  Plotted are the fitted center velocity,
$V_{\rm LSR}$, the full-width at half-maximum (FWHM) line width,
$\Delta V$, and their associated errors.  The $V_{\rm LSR}$ and
$\Delta V$ images are similar to first and second moment maps,
respectively, but in detail they are different mathematical
operations.  Figures~\ref{fig:grad-atca}--\ref{fig:grad-vla} reveal
clear velocity gradients across the roughly circular nebulae.
G297.651$-$00.973 and G035.126$-$0.755 illustrate particularly good
examples, but we detect velocity gradients in 44\% and 25\% of \hii\
regions in the ATCA and VLA surveys, respectively.  Less common,
however, is the centrally-peaked distribution of the FWHM line width
detected in G297.651$-$00.973 and G035.126$-$0.755.

To characterize the velocity gradients in our \hii\ region sample, we
fit both a constant model (no velocity gradient) and a plane model
(velocity gradient) to the $V_{\rm LSR}$ image using maximum
likelihood estimation (MLE)---see Appendix~\ref{sec:mle} for details.
For each model, we calculate the Bayesian Information Criterion (BIC).
The BIC is used to select the best model that does not overfit the
data.  If
$\Delta{\rm BIC} \equiv {\rm BIC(constant)} - \rm{BIC(plane)} \ge 6$
then the plane model is strongly preferred \citep{kass95}.  A total of
212 and 19 detections are best fit by a plane model for the ATCA and VLA,
respectively.  Visual inspection of both the Gaussian fit images and
the data cubes for all sources with $\Delta{\rm BIC} \ge 6$ reveals
problems with the velocity gradient fit for 46 and 7 detections for the
ATCA and VLA, respectively.  For example, the velocity gradient was
either fit across two distinct sources or the source was cutoff by the
primary beam.  Removing these sources leaves a total of 166 and 12
nebulae with bona fide velocity gradients for the ATCA and VLA,
respectively.  Finally, to provide a quantitative measure of the
goodness of fit we calculate the root mean square error (RMSE) between
the model and data.

Tables~\ref{tab:atca} and \ref{tab:vla} summarize the properties of
our velocity gradient \hii\ regions for the ATCA and VLA,
respectively.  Listed are the source name, the source field, the MLE
velocity gradient fit parameters, the quality factor, QF, the number
of synthesized beams across the nebula, the RRL peak SNR, and the
RMSE.  The velocity gradient fit parameters consist of the velocity
offset, $c_{0}$, the velocity gradient magnitude,
$\nabla V_{\rm LSR}$, and the position angle, PA. The ATCA primary
beam is large and often contains multiple \hii\ regions.  Therefore,
we have observed some sources multiple times in the SHRDS and these
duplicates, consisting of 16 \hii\ regions, are included in
Table~\ref{tab:atca} because they are independent measurements.  They
can be distinguished by the different field names.

\begin{figure}
  \centering
  \includegraphics[angle=0,scale=0.8]{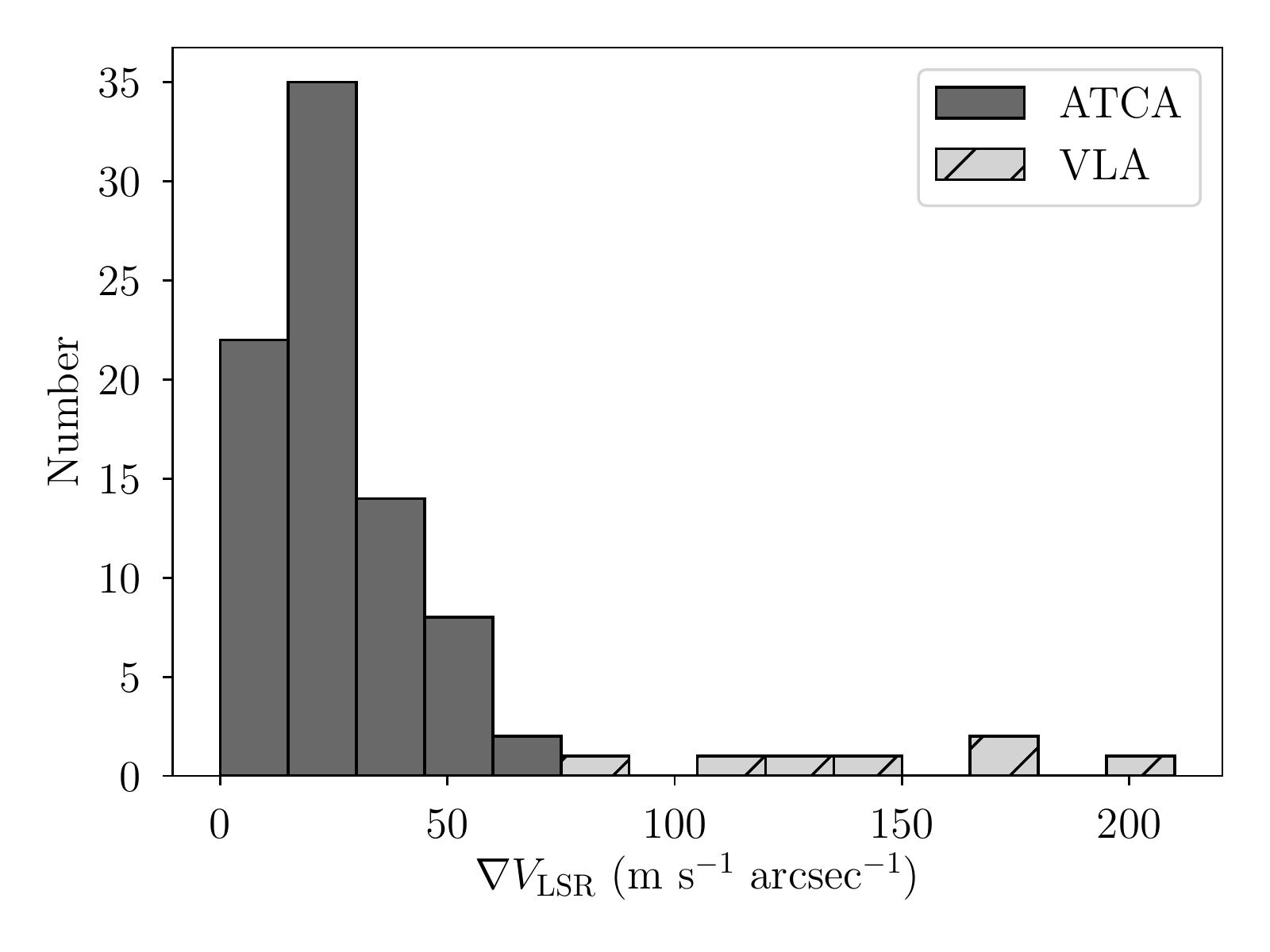}
  \includegraphics[angle=0,scale=0.8]{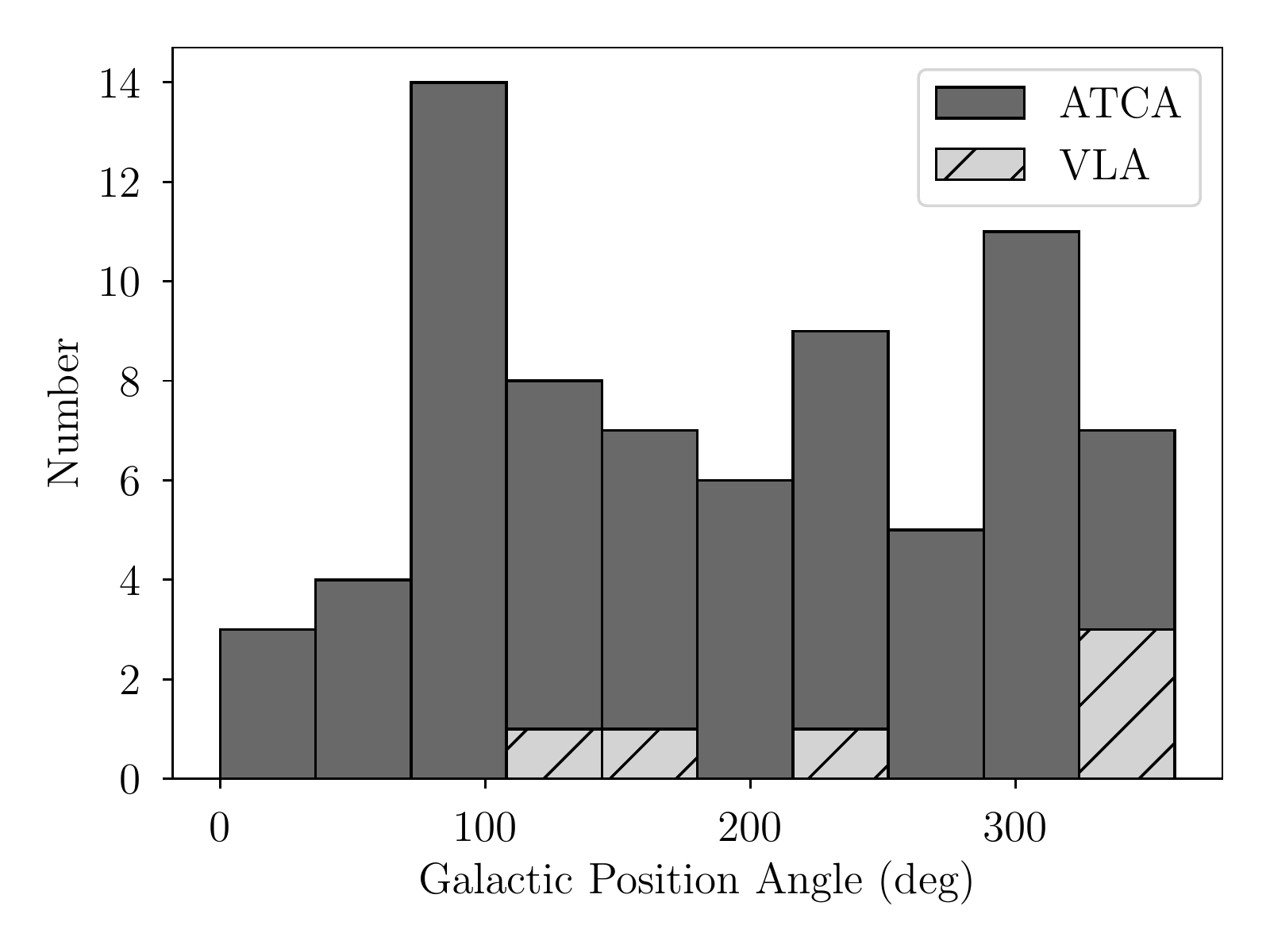}
  \caption{Histograms of the velocity gradient magnitude,
    $\nabla V_{\rm LSR}$, (top) and the position angle, PA, (bottom)
    for the ATCA and VLA \hii\ region samples. Only sources with RMSE
    less than 1.0\kms\ are included.}
\label{fig:grad-fits}
\end{figure}

The distribution of velocity gradient magnitudes and position angles
are shown in Figure~\ref{fig:grad-fits}.  Only unique sources with
RMSE values less than 1.0\kms\ are included.  Based on visual
inspection of the $V_{\rm LSR}$ images and the velocity gradient fits,
we deem that sources with RMSE less than 1.0\kms\ are well
characterized by velocity gradients.  Using this threshold for the
RMSE excludes 74 and 5 detections from the ATCA and VLA \hii\ region
sample, respectively.  The ATCA \hii\ region sample velocity gradient
magnitudes range between $\nabla V_{\rm LSR} = 5-80$\msarcsec, whereas
the VLA \hii\ region sample have higher values,
$\nabla V_{\rm LSR} = 80-200$\msarcsec.  Since the VLA spatial
resolution is about 10 times better than the ATCA we are not
necessarily comparing the same area of the \hii\ region; this, of
course, depends on the distance.  Calculating the velocity gradient in
physical units, \kmspc, may provide more stringent constraints on
dynamical models, but here we decide to use the distance independent
units of \msarcsec\ since we do not yet have accurate distances to
many of our sources.  The number of nebulae deceases with increasing
velocity gradient magnitude and turns over near
$\nabla V_{\rm LSR} = 15$\msarcsec.  There are likely \hii\ regions
with velocity gradients less than this value that we do not detect due
to sensitivity.

\clearpage

\startlongtable 
\begin{deluxetable}{llrcccccc} 
  \tabletypesize{\scriptsize} 
  \tablecaption{ATCA \hii\ Region Sample}\label{tab:atca} 
  \tablewidth{0pt} 
  \setlength{\tabcolsep}{2.0pt} 
  \tablehead{\colhead{} & \colhead{} & 
    \multicolumn{3}{c}{\underline{~~~~~~~~~~~MLE Velocity Gradient Fit~~~~~~~~~~~}} & 
    \colhead{} & \colhead{} & \colhead{} & \colhead{} \\ 
    \colhead{} & \colhead{} & 
    \colhead{$c_{0}$} & \colhead{$\nabla V_{\rm LSR}$} & \colhead{PA} & 
    \colhead{} & \colhead{No.} & \colhead{Peak} & \colhead{RMSE} \\ 
    \colhead{Source} & \colhead{Field} & 
    \colhead{(\kms)} & \colhead{(\msarcsec)} & \colhead{(degree)} & 
    \colhead{QF} & \colhead{Beams} & \colhead{SNR} & \colhead{(\kms)} } 
  \startdata 
G012.804$-$00.207 & g012.804$-$       & $35.65 \pm 0.07$          & $8.53 \pm 0.84$           & $265.26 \pm 7.34$         & A & 3.6 & 281.00  & 0.74  \\ 
G013.880+00.285   & overlap4          & $50.76 \pm 0.04$          & $14.94 \pm 0.70$          & $179.79 \pm 3.02$         & A & 3.8 & 449.70  & 0.58  \\ 
G233.753$-$00.193 & ch1               & $37.45 \pm 0.30$          & $41.93 \pm 3.92$          & $89.01 \pm 6.08$          & A & 2.1 & 20.10   & 1.86  \\ 
G259.057$-$01.544 & shrds030          & $59.18 \pm 0.51$          & $317.62 \pm 18.12$        & $126.50 \pm 3.12$         & A & 1.0 & 25.40   & 9.12  \\ 
G263.615$-$00.534 & ch5               & $0.08 \pm 0.08$           & $21.73 \pm 2.06$          & $155.71 \pm 5.53$         & A & 5.3 & 91.80   & 1.03  \\ 
G264.343+01.457   & ch6               & $13.27 \pm 0.09$          & $36.50 \pm 2.64$          & $40.66 \pm 4.07$          & A & 4.1 & 98.00   & 0.69  \\ 
G265.151+01.454   & ch7               & $6.03 \pm 0.02$           & $71.23 \pm 0.73$          & $74.12 \pm 0.53$          & A & 8.3 & 218.20  & 4.21  \\ 
G281.175$-$01.645 & ch17              & $-6.98 \pm 0.13$          & $14.54 \pm 3.69$          & $49.56 \pm 14.21$         & A & 3.3 & 87.60   & 0.59  \\ 
G281.175$-$01.645 & caswell1          & $-7.43 \pm 0.18$          & $15.74 \pm 4.16$          & $19.90 \pm 15.79$         & A & 2.8 & 58.90   & 0.64  \\ 
G282.015$-$00.997 & shrds1007         & $0.29 \pm 0.13$           & $26.61 \pm 1.76$          & $124.47 \pm 3.63$         & B & 3.8 & 40.60   & 0.82  \\ 
G282.027$-$01.182 & ch19              & $21.20 \pm 0.06$          & $31.29 \pm 1.15$          & $335.36 \pm 2.35$         & A & 6.8 & 286.60  & 3.34  \\ 
G282.842$-$01.252 & shrds1232         & $-4.61 \pm 0.23$          & $16.16 \pm 4.75$          & $281.33 \pm 17.58$        & A & 2.4 & 36.50   & 1.17  \\ 
G284.712+00.317   & caswell2          & $9.26 \pm 0.12$           & $39.98 \pm 2.51$          & $237.68 \pm 3.64$         & A & 4.5 & 83.80   & 1.05  \\ 
G284.712+00.317   & ch32              & $10.91 \pm 0.11$          & $48.64 \pm 2.28$          & $242.35 \pm 2.85$         & A & 5.6 & 118.30  & 1.38  \\ 
G285.260$-$00.051 & ch33              & $-1.02 \pm 0.06$          & $26.37 \pm 1.11$          & $177.09 \pm 1.62$         & A & 7.6 & 229.70  & 1.50  \\ 
G286.362$-$00.297 & shrds1017         & $-33.68 \pm 0.19$         & $52.57 \pm 3.25$          & $190.91 \pm 3.98$         & B & 3.3 & 34.20   & 0.83  \\ 
G286.362$-$00.297 & shrds1018g        & $-39.76 \pm 0.27$         & $49.51 \pm 3.16$          & $200.64 \pm 2.79$         & B & 1.9 & 17.70   & 0.48  \\ 
G286.391$-$01.351 & ch35              & $38.27 \pm 0.18$          & $34.71 \pm 5.35$          & $5.18 \pm 6.16$           & A & 2.6 & 64.10   & 0.79  \\ 
G291.046$-$02.079 & shrds199          & $-18.93 \pm 0.19$         & $37.70 \pm 3.68$          & $227.54 \pm 5.59$         & A & 3.1 & 40.20   & 0.48  \\ 
G291.281$-$00.726 & caswell3          & $-23.38 \pm 0.07$         & $46.15 \pm 1.39$          & $159.23 \pm 1.65$         & A & 4.9 & 226.70  & 0.86  \\ 
G291.281$-$00.726 & ch52              & $-23.58 \pm 0.06$         & $64.87 \pm 1.29$          & $163.43 \pm 0.89$         & A & 6.3 & 407.40  & 1.24  \\ 
G291.863$-$00.682 & ch55              & $24.92 \pm 0.08$          & $12.92 \pm 1.78$          & $329.46 \pm 9.20$         & A & 4.5 & 144.60  & 0.91  \\ 
G291.863$-$00.682 & caswell4          & $24.31 \pm 0.10$          & $9.71 \pm 2.09$           & $317.15 \pm 12.31$        & A & 3.9 & 122.60  & 0.54  \\ 
G293.024$-$01.029 & ch57              & $65.77 \pm 0.15$          & $19.83 \pm 3.98$          & $132.85 \pm 11.51$        & A & 3.8 & 65.40   & 0.68  \\ 
G293.967$-$00.984 & shrds219          & $28.46 \pm 0.50$          & $36.56 \pm 12.10$         & $83.34 \pm 16.46$         & B & 1.1 & 28.20   & 0.38  \\ 
G297.651$-$00.973 & ch65              & $31.76 \pm 0.22$          & $67.50 \pm 5.91$          & $94.55 \pm 5.09$          & A & 2.9 & 46.50   & 0.71  \\ 
G298.183$-$00.784 & caswell5          & $17.97 \pm 0.12$          & $13.18 \pm 3.05$          & $323.02 \pm 13.28$        & A & 3.1 & 131.80  & 0.66  \\ 
G298.183$-$00.784 & ch66              & $17.63 \pm 0.12$          & $11.14 \pm 3.40$          & $280.26 \pm 18.76$        & A & 3.1 & 165.50  & 0.93  \\ 
G298.224$-$00.334 & ch67              & $31.79 \pm 0.07$          & $60.01 \pm 1.18$          & $36.16 \pm 1.25$          & A & 5.6 & 349.80  & 3.06  \\ 
G298.846+00.121   & shrds258          & $19.65 \pm 0.29$          & $29.72 \pm 5.80$          & $111.24 \pm 10.90$        & A & 1.5 & 38.60   & 0.47  \\ 
G299.349$-$00.267 & ch71              & $-39.61 \pm 0.20$         & $16.16 \pm 2.85$          & $350.31 \pm 8.49$         & B & 2.4 & 38.00   & 0.54  \\ 
G299.349$-$00.267 & caswell6          & $-40.35 \pm 0.18$         & $24.88 \pm 3.62$          & $331.40 \pm 8.18$         & A & 3.0 & 38.80   & 0.42  \\ 
G300.965+01.162   & g300.983+01.117   & $-40.93 \pm 0.16$         & $30.18 \pm 1.13$          & $186.87 \pm 4.41$         & B & 3.4 & 49.00   & 1.54  \\ 
G301.116+00.968   & caswell8          & $-40.76 \pm 0.08$         & $41.28 \pm 1.62$          & $217.31 \pm 2.22$         & A & 4.2 & 171.30  & 0.51  \\ 
G301.116+00.968   & ch74              & $-40.31 \pm 0.07$         & $57.89 \pm 1.76$          & $215.77 \pm 1.65$         & A & 4.7 & 228.90  & 0.74  \\ 
G302.436$-$00.106 & shrds294          & $-40.37 \pm 0.35$         & $25.44 \pm 7.25$          & $93.56 \pm 18.51$         & A & 2.0 & 25.10   & 0.51  \\ 
G302.636$-$00.672 & shrds297          & $30.90 \pm 0.21$          & $11.81 \pm 3.98$          & $94.36 \pm 18.10$         & A & 2.4 & 47.00   & 0.62  \\ 
G302.805+01.287   & ch79              & $-28.44 \pm 0.19$         & $22.78 \pm 2.47$          & $76.57 \pm 4.52$          & A & 2.8 & 54.50   & 0.41  \\ 
G303.342$-$00.718 & shrds303          & $27.65 \pm 0.48$          & $35.41 \pm 12.57$         & $75.86 \pm 18.23$         & B & 1.1 & 29.00   & 0.56  \\ 
G304.465$-$00.023 & shrds336          & $-15.37 \pm 0.11$         & $23.04 \pm 1.99$          & $109.09 \pm 2.59$         & A & 3.6 & 72.00   & 0.52  \\ 
G305.362+00.197   & ch86              & $-36.88 \pm 0.06$         & $25.43 \pm 0.75$          & $203.99 \pm 2.36$         & A & 5.5 & 350.30  & 2.29  \\ 
G306.321$-$00.358 & ch93              & $-19.54 \pm 0.11$         & $33.87 \pm 1.88$          & $16.27 \pm 4.01$          & A & 4.6 & 86.20   & 0.84  \\ 
G308.916+00.124   & shrds384          & $-43.45 \pm 0.33$         & $53.30 \pm 9.24$          & $124.50 \pm 9.85$         & A & 1.9 & 45.90   & 1.43  \\ 
G309.151$-$00.215 & shrds385          & $-10.64 \pm 0.13$         & $27.69 \pm 1.78$          & $274.02 \pm 4.36$         & B & 4.6 & 33.20   & 1.01  \\ 
G310.260$-$00.199 & shrds403          & $7.21 \pm 0.31$           & $172.74 \pm 7.18$         & $54.07 \pm 1.91$          & B & 1.6 & 18.50   & 3.85  \\ 
G310.519$-$00.220 & shrds407          & $23.53 \pm 0.37$          & $41.80 \pm 11.63$         & $290.60 \pm 14.95$        & A & 1.4 & 35.30   & 0.33  \\ 
G311.563+00.239   & shrds428          & $-4.73 \pm 0.23$          & $226.11 \pm 5.38$         & $156.74 \pm 1.71$         & A & 1.8 & 30.10   & 8.57  \\ 
G311.629+00.289   & caswell12         & $-54.95 \pm 0.10$         & $21.76 \pm 2.23$          & $326.00 \pm 5.92$         & A & 3.4 & 179.40  & 1.08  \\ 
G311.629+00.289   & ch110             & $-58.26 \pm 0.11$         & $41.84 \pm 2.19$          & $312.46 \pm 2.85$         & A & 3.4 & 176.70  & 1.56  \\ 
G311.809$-$00.309 & shrds1103         & $-37.23 \pm 0.17$         & $51.09 \pm 3.26$          & $235.14 \pm 4.21$         & A & 4.1 & 40.80   & 1.13  \\ 
G311.893+00.086   & ch113             & $-47.90 \pm 0.04$         & $29.08 \pm 0.80$          & $283.14 \pm 1.41$         & A & 6.0 & 160.70  & 1.81  \\ 
G311.919+00.204   & ch114             & $-43.26 \pm 0.06$         & $46.34 \pm 0.94$          & $346.30 \pm 1.42$         & A & 6.6 & 71.40   & 1.79  \\ 
G311.963$-$00.037 & shrds439          & $-54.90 \pm 0.12$         & $8.63 \pm 1.92$           & $52.44 \pm 12.96$         & A & 3.8 & 46.40   & 0.49  \\ 
G313.671$-$00.105 & atca348           & $-53.50 \pm 0.42$         & $30.03 \pm 9.44$          & $278.44 \pm 18.07$        & A & 1.2 & 21.10   & 0.53  \\ 
G313.790+00.705   & atca352           & $-58.04 \pm 0.25$         & $34.73 \pm 4.92$          & $162.90 \pm 8.83$         & A & 2.1 & 33.50   & 0.65  \\ 
G315.312$-$00.272 & ch121             & $15.41 \pm 0.29$          & $15.77 \pm 4.35$          & $7.60 \pm 18.34$          & A & 1.7 & 25.90   & 0.55  \\ 
G315.312$-$00.272 & ch87.1            & $14.83 \pm 0.23$          & $30.89 \pm 4.21$          & $6.15 \pm 8.81$           & A & 2.4 & 33.50   & 0.28  \\ 
G316.796$-$00.056 & ch124             & $-36.72 \pm 0.05$         & $33.45 \pm 0.66$          & $300.23 \pm 1.17$         & A & 6.1 & 288.80  & 2.04  \\ 
G317.405+00.091   & shrds1122         & $-39.31 \pm 0.08$         & $14.82 \pm 1.52$          & $96.61 \pm 6.16$          & A & 3.0 & 126.90  & 0.70  \\ 
G317.861+00.160   & atca402           & $1.65 \pm 0.18$           & $12.15 \pm 3.77$          & $212.19 \pm 15.85$        & A & 3.0 & 45.00   & 0.96  \\ 
G318.233$-$00.605 & shrds1126         & $-39.01 \pm 0.11$         & $83.88 \pm 1.84$          & $314.85 \pm 1.25$         & B & 5.7 & 29.10   & 3.74  \\ 
G318.915$-$00.165 & ch131             & $-29.40 \pm 0.11$         & $9.34 \pm 2.80$           & $79.76 \pm 10.35$         & A & 3.3 & 192.80  & 0.56  \\ 
G319.188$-$00.329 & shrds1129         & $-27.10 \pm 0.11$         & $35.44 \pm 1.46$          & $327.19 \pm 2.21$         & B & 6.9 & 39.30   & 1.89  \\ 
G319.229+00.225   & atca412           & $-66.18 \pm 0.30$         & $20.36 \pm 4.12$          & $191.18 \pm 13.41$        & A & 2.1 & 21.80   & 0.40  \\ 
G319.884+00.793   & ch134             & $-43.14 \pm 0.09$         & $23.86 \pm 1.02$          & $0.47 \pm 4.59$           & A & 4.6 & 75.40   & 0.62  \\ 
G320.163+00.797   & ch136             & $-44.24 \pm 0.06$         & $74.05 \pm 0.67$          & $29.55 \pm 0.70$          & A & 7.1 & 176.80  & 0.64  \\ 
G320.236$-$00.099 & shrds1140         & $-6.73 \pm 0.09$          & $23.01 \pm 1.63$          & $298.49 \pm 4.34$         & B & 4.4 & 57.90   & 1.03  \\ 
G320.884$-$00.641 & shrds561          & $7.88 \pm 0.26$           & $49.17 \pm 5.77$          & $357.57 \pm 6.12$         & A & 2.4 & 27.50   & 0.34  \\ 
G321.725+01.169   & caswell14         & $-32.34 \pm 0.07$         & $4.86 \pm 1.53$           & $220.55 \pm 18.15$        & A & 3.5 & 168.90  & 0.57  \\ 
G322.162+00.625   & caswell15         & $-52.28 \pm 0.05$         & $20.29 \pm 0.89$          & $110.15 \pm 2.40$         & A & 3.6 & 336.00  & 0.61  \\ 
G322.162+00.625   & ch145             & $-51.80 \pm 0.04$         & $29.99 \pm 0.78$          & $116.38 \pm 1.16$         & A & 4.6 & 550.80  & 1.19  \\ 
G323.806+00.020   & atca459           & $-59.36 \pm 0.23$         & $17.41 \pm 3.91$          & $231.87 \pm 14.09$        & A & 2.5 & 36.40   & 0.25  \\ 
G324.201+00.117   & caswell16         & $-90.15 \pm 0.08$         & $6.80 \pm 1.71$           & $152.72 \pm 13.73$        & A & 3.5 & 191.60  & 0.75  \\ 
G324.642$-$00.321 & atca466           & $-49.65 \pm 0.29$         & $22.53 \pm 4.87$          & $70.29 \pm 14.78$         & A & 1.8 & 32.50   & 0.46  \\ 
G325.108+00.053   & atca472           & $-66.58 \pm 0.32$         & $45.74 \pm 3.08$          & $65.18 \pm 3.67$          & B & 2.0 & 20.90   & 0.97  \\ 
G325.354$-$00.036 & atca475           & $-65.43 \pm 0.44$         & $24.84 \pm 6.94$          & $175.12 \pm 22.15$        & A & 1.2 & 21.20   & 0.63  \\ 
G326.446+00.901   & caswell17         & $-40.35 \pm 0.05$         & $23.45 \pm 1.02$          & $226.91 \pm 2.48$         & A & 4.1 & 226.20  & 1.09  \\ 
G326.446+00.901   & ch154             & $-39.72 \pm 0.05$         & $28.77 \pm 1.02$          & $226.92 \pm 2.07$         & A & 4.7 & 264.30  & 1.49  \\ 
G326.473$-$00.378 & shrds609          & $-54.87 \pm 0.32$         & $41.56 \pm 8.94$          & $239.43 \pm 12.70$        & A & 1.8 & 42.70   & 1.13  \\ 
G326.721+00.773   & atca484           & $-40.57 \pm 0.17$         & $30.72 \pm 3.46$          & $294.96 \pm 5.76$         & A & 2.6 & 34.30   & 0.33  \\ 
G327.714+00.576   & atca498           & $-49.22 \pm 0.40$         & $39.29 \pm 6.44$          & $66.91 \pm 9.46$          & A & 1.7 & 23.80   & 0.44  \\ 
G328.117+00.108   & shrds647          & $-103.02 \pm 0.16$        & $23.51 \pm 2.89$          & $87.93 \pm 7.92$          & A & 3.2 & 38.80   & 0.84  \\ 
G328.945+00.570   & shrds669          & $-91.67 \pm 0.16$         & $48.77 \pm 3.56$          & $210.49 \pm 4.17$         & A & 2.9 & 52.90   & 1.02  \\ 
G329.266+00.111   & shrds677          & $-73.83 \pm 0.36$         & $48.76 \pm 8.74$          & $343.27 \pm 8.65$         & A & 1.5 & 16.10   & 0.28  \\ 
G330.039$-$00.058 & ch174             & $-40.49 \pm 0.18$         & $11.72 \pm 2.08$          & $62.31 \pm 9.87$          & A & 2.8 & 51.60   & 0.72  \\ 
G330.673$-$00.388 & shrds1171         & $-64.06 \pm 0.10$         & $11.13 \pm 2.67$          & $195.04 \pm 4.45$         & B & 3.6 & 100.50  & 1.01  \\ 
G330.738$-$00.449 & shrds1171         & $-61.23 \pm 0.20$         & $21.16 \pm 5.72$          & $339.52 \pm 16.28$        & A & 2.8 & 36.90   & 1.47  \\ 
G330.873$-$00.369 & ch177             & $-53.37 \pm 0.06$         & $9.31 \pm 1.09$           & $236.85 \pm 6.17$         & A & 5.5 & 204.90  & 2.09  \\ 
G331.123$-$00.530 & shrds717          & $-68.06 \pm 0.05$         & $20.34 \pm 0.81$          & $283.67 \pm 2.04$         & A & 4.4 & 157.70  & 1.48  \\ 
G331.145+00.133   & shrds719          & $-75.94 \pm 0.31$         & $44.57 \pm 4.21$          & $332.45 \pm 6.91$         & A & 2.1 & 31.00   & 1.28  \\ 
G331.156$-$00.391 & shrds720          & $-62.25 \pm 0.11$         & $27.73 \pm 2.10$          & $232.81 \pm 4.19$         & A & 3.5 & 55.30   & 1.33  \\ 
G331.172$-$00.460 & shrds721          & $-71.10 \pm 0.14$         & $76.61 \pm 3.73$          & $236.03 \pm 2.71$         & A & 2.6 & 69.40   & 3.67  \\ 
G331.249+01.071   & shrds1174         & $-82.31 \pm 0.15$         & $35.80 \pm 2.00$          & $247.14 \pm 3.61$         & A & 4.8 & 40.00   & 1.08  \\ 
G331.259$-$00.188 & ch180             & $-83.60 \pm 0.07$         & $37.30 \pm 1.13$          & $21.59 \pm 1.59$          & A & 6.0 & 133.40  & 1.58  \\ 
G331.361$-$00.019 & ch182             & $-83.09 \pm 0.05$         & $31.35 \pm 0.92$          & $98.47 \pm 1.76$          & B & 6.2 & 157.30  & 1.40  \\ 
G331.468$-$00.262 & shrds732          & $-100.51 \pm 0.19$        & $19.85 \pm 4.15$          & $278.40 \pm 15.68$        & A & 1.8 & 31.20   & 0.42  \\ 
G331.653+00.128   & shrds741          & $-90.28 \pm 0.07$         & $24.90 \pm 1.36$          & $132.31 \pm 3.18$         & A & 4.6 & 97.40   & 0.87  \\ 
G331.744$-$00.068 & shrds743          & $-88.64 \pm 0.28$         & $29.93 \pm 8.30$          & $81.58 \pm 16.16$         & A & 1.8 & 37.70   & 0.45  \\ 
G331.834$-$00.002 & shrds746          & $-80.59 \pm 0.18$         & $32.09 \pm 3.81$          & $111.28 \pm 6.69$         & A & 3.8 & 41.80   & 1.03  \\ 
G332.311$-$00.567 & shrds759          & $-55.73 \pm 0.14$         & $25.76 \pm 3.36$          & $269.05 \pm 7.51$         & A & 2.5 & 56.90   & 0.49  \\ 
G332.957+01.793   & shrds782          & $-26.71 \pm 0.36$         & $100.45 \pm 6.59$         & $317.87 \pm 3.82$         & B & 2.6 & 17.70   & 1.55  \\ 
G332.964+00.771   & ch191             & $-51.19 \pm 0.04$         & $25.60 \pm 0.59$          & $319.69 \pm 1.35$         & B & 8.8 & 182.60  & 2.05  \\ 
G332.987+00.902   & shrds784          & $-45.46 \pm 0.16$         & $23.76 \pm 4.04$          & $211.23 \pm 9.75$         & A & 3.0 & 39.60   & 0.45  \\ 
G332.990$-$00.619 & shrds785          & $-49.78 \pm 0.06$         & $63.63 \pm 1.47$          & $93.38 \pm 0.91$          & A & 6.4 & 89.30   & 1.51  \\ 
G333.052+00.033   & shrds788          & $-38.94 \pm 0.31$         & $55.96 \pm 5.01$          & $0.84 \pm 6.68$           & A & 2.1 & 33.70   & 2.09  \\ 
G333.129$-$00.439 & shrds789          & $-28.34 \pm 0.05$         & $83.51 \pm 0.23$          & $183.36 \pm 1.12$         & B & 2.2 & 262.90  & 4.55  \\ 
G333.164$-$00.100 & shrds792          & $-90.03 \pm 0.08$         & $27.03 \pm 0.66$          & $178.78 \pm 5.36$         & B & 3.6 & 71.30   & 1.06  \\ 
G333.255+00.065   & shrds794          & $-54.07 \pm 0.10$         & $8.70 \pm 1.72$           & $235.65 \pm 11.87$        & A & 3.6 & 126.60  & 2.77  \\ 
G333.467$-$00.159 & shrds799          & $-42.18 \pm 0.08$         & $12.38 \pm 2.26$          & $346.70 \pm 10.83$        & A & 2.4 & 129.40  & 0.29  \\ 
G333.534$-$00.383 & shrds801          & $-55.93 \pm 0.09$         & $76.15 \pm 1.81$          & $35.11 \pm 1.13$          & B & 5.9 & 46.30   & 1.12  \\ 
G334.202+00.193   & shrds811          & $-95.70 \pm 0.34$         & $16.79 \pm 6.43$          & $217.96 \pm 19.73$        & A & 1.6 & 30.80   & 0.81  \\ 
G334.721$-$00.653 & caswell20         & $14.86 \pm 0.27$          & $45.16 \pm 7.35$          & $323.07 \pm 9.29$         & A & 2.0 & 58.10   & 0.32  \\ 
G334.721$-$00.653 & ch201             & $11.88 \pm 0.27$          & $46.41 \pm 5.29$          & $333.75 \pm 9.31$         & A & 2.0 & 66.10   & 0.56  \\ 
G334.774$-$00.023 & shrds828          & $-26.41 \pm 0.24$         & $18.43 \pm 5.67$          & $313.78 \pm 17.56$        & A & 2.7 & 42.50   & 0.58  \\ 
G336.026$-$00.817 & shrds843          & $-45.78 \pm 0.16$         & $14.15 \pm 3.79$          & $255.51 \pm 16.60$        & A & 2.3 & 58.90   & 0.30  \\ 
G336.086$-$00.074 & shrds845          & $-97.24 \pm 0.30$         & $55.06 \pm 6.35$          & $334.25 \pm 7.17$         & B & 1.9 & 16.40   & 0.89  \\ 
G336.367$-$00.003 & shrds852          & $-129.13 \pm 0.23$        & $20.63 \pm 4.02$          & $343.21 \pm 10.21$        & A & 1.9 & 52.70   & 0.47  \\ 
G336.491$-$01.474 & ch209             & $-23.13 \pm 0.00$         & $7.42 \pm 0.05$           & $86.89 \pm 1.55$          & A & 4.4 & 340.60  & 0.95  \\ 
G336.491$-$01.474 & caswell21         & $-23.59 \pm 0.06$         & $10.55 \pm 1.25$          & $263.69 \pm 6.83$         & A & 3.7 & 304.30  & 0.51  \\ 
G337.286+00.113   & shrds871          & $-108.53 \pm 0.16$        & $20.80 \pm 3.06$          & $125.06 \pm 8.76$         & A & 2.5 & 50.10   & 0.50  \\ 
G337.496$-$00.258 & ch214             & $-94.39 \pm 0.15$         & $54.46 \pm 2.06$          & $71.84 \pm 1.81$          & B & 5.3 & 36.80   & 1.61  \\ 
G337.665$-$00.176 & shrds883          & $-50.66 \pm 0.20$         & $38.95 \pm 4.57$          & $279.45 \pm 6.53$         & A & 2.8 & 65.30   & 2.14  \\ 
G337.684$-$00.343 & shrds884          & $-42.99 \pm 0.09$         & $38.96 \pm 1.77$          & $110.19 \pm 2.84$         & A & 5.5 & 75.50   & 1.47  \\ 
G337.705$-$00.059 & shrds885          & $-48.61 \pm 0.18$         & $17.42 \pm 2.99$          & $45.14 \pm 9.85$          & A & 2.0 & 27.40   & 1.01  \\ 
G337.922$-$00.463 & ch216             & $-38.13 \pm 0.04$         & $34.05 \pm 0.43$          & $117.91 \pm 0.89$         & B & 8.2 & 332.60  & 4.61  \\ 
G338.405$-$00.203 & caswell22         & $0.44 \pm 0.11$           & $20.63 \pm 2.78$          & $95.61 \pm 7.46$          & A & 3.4 & 160.10  & 0.65  \\ 
G338.462$-$00.262 & ch221             & $-52.55 \pm 0.20$         & $31.94 \pm 1.29$          & $247.86 \pm 3.74$         & B & 2.5 & 37.70   & 1.67  \\ 
G338.565$-$00.151 & shrds899          & $-114.31 \pm 0.18$        & $15.50 \pm 2.58$          & $194.05 \pm 8.88$         & A & 2.4 & 56.90   & 0.77  \\ 
G338.576+00.020   & shrds900          & $-23.66 \pm 0.08$         & $18.66 \pm 1.56$          & $206.82 \pm 3.50$         & B & 3.4 & 101.00  & 0.69  \\ 
G338.837$-$00.318 & shrds904          & $-123.39 \pm 0.27$        & $30.74 \pm 5.43$          & $47.62 \pm 10.45$         & A & 1.8 & 29.80   & 0.42  \\ 
G338.883+00.537   & shrds1225         & $-60.79 \pm 0.23$         & $80.43 \pm 7.38$          & $192.10 \pm 5.57$         & A & 1.0 & 35.60   & 0.83  \\ 
G338.934$-$00.067 & ch225             & $-39.57 \pm 0.06$         & $66.12 \pm 0.79$          & $188.55 \pm 0.74$         & A & 7.9 & 74.40   & 1.39  \\ 
G339.275$-$00.607 & shrds915          & $-39.01 \pm 0.24$         & $26.38 \pm 3.89$          & $209.72 \pm 8.69$         & B & 2.2 & 30.00   & 0.58  \\ 
G339.478+00.181   & shrds917          & $-84.06 \pm 0.22$         & $21.98 \pm 4.37$          & $249.79 \pm 12.60$        & A & 2.0 & 55.90   & 0.41  \\ 
G339.845+00.299   & ch230             & $-15.51 \pm 0.11$         & $46.02 \pm 1.41$          & $168.30 \pm 2.40$         & A & 4.4 & 76.00   & 1.15  \\ 
G339.952+00.052   & shrds1234         & $-121.43 \pm 0.28$        & $56.92 \pm 9.29$          & $325.86 \pm 9.91$         & A & 1.6 & 31.10   & 0.94  \\ 
G340.294$-$00.193 & ch234             & $-42.06 \pm 0.04$         & $29.71 \pm 0.46$          & $182.79 \pm 1.24$         & B & 8.4 & 118.50  & 0.87  \\ 
G342.062+00.417   & ch239             & $-71.01 \pm 0.07$         & $43.75 \pm 1.16$          & $252.40 \pm 1.27$         & B & 6.4 & 163.10  & 1.56  \\ 
G342.062+00.417   & caswell23         & $-67.99 \pm 0.07$         & $41.88 \pm 1.57$          & $256.22 \pm 2.02$         & A & 4.7 & 149.50  & 0.88  \\ 
G343.914$-$00.646 & caswell24         & $-26.80 \pm 0.11$         & $17.13 \pm 1.82$          & $121.97 \pm 5.97$         & B & 7.6 & 33.10   & 0.65  \\ 
G344.424+00.044   & ch246             & $-63.20 \pm 0.06$         & $10.34 \pm 1.21$          & $35.98 \pm 7.50$          & A & 3.9 & 314.20  & 0.60  \\ 
G344.424+00.044   & caswell25         & $-63.60 \pm 0.06$         & $6.38 \pm 1.52$           & $49.53 \pm 13.63$         & A & 3.5 & 248.60  & 0.35  \\ 
G345.391+01.398   & ch251             & $-14.45 \pm 0.03$         & $1.81 \pm 0.55$           & $127.03 \pm 16.14$        & B & 7.9 & 241.60  & 1.35  \\ 
G345.410$-$00.953 & ch252             & $-20.09 \pm 0.04$         & $44.97 \pm 0.46$          & $7.53 \pm 0.74$           & A & 6.9 & 436.30  & 1.57  \\ 
G345.432+00.207   & ch253             & $-7.87 \pm 0.20$          & $17.74 \pm 3.84$          & $192.02 \pm 14.95$        & A & 3.1 & 62.70   & 0.96  \\ 
G345.651+00.015   & ch256             & $-10.06 \pm 0.09$         & $13.54 \pm 2.10$          & $350.04 \pm 9.20$         & A & 4.1 & 260.70  & 0.70  \\ 
G346.521+00.086   & ch260             & $5.57 \pm 0.18$           & $9.70 \pm 2.97$           & $252.18 \pm 12.13$        & A & 2.6 & 86.20   & 0.88  \\ 
G348.000$-$00.496 & ch264             & $-95.98 \pm 0.08$         & $4.79 \pm 0.92$           & $58.06 \pm 10.96$         & B & 7.4 & 70.10   & 1.77  \\ 
G348.249$-$00.971 & ch266             & $-18.80 \pm 0.05$         & $33.86 \pm 0.83$          & $38.98 \pm 1.39$          & A & 8.5 & 345.00  & 2.10  \\ 
G348.710$-$01.044 & ch268             & $-12.23 \pm 0.03$         & $25.53 \pm 0.49$          & $352.05 \pm 0.98$         & B & 8.2 & 381.00  & 1.35  \\ 
G350.105+00.081   & ch273             & $-70.20 \pm 0.04$         & $34.30 \pm 0.74$          & $285.60 \pm 0.83$         & B & 6.5 & 336.00  & 1.12  \\ 
G350.505+00.956   & ch274             & $-9.63 \pm 0.08$          & $14.95 \pm 1.79$          & $110.67 \pm 4.54$         & A & 4.0 & 281.90  & 1.06  \\ 
G351.246+00.673   & overlap1          & $0.21 \pm 0.05$           & $25.67 \pm 0.95$          & $212.80 \pm 2.38$         & A & 3.4 & 342.20  & 1.19  \\ 
G351.472$-$00.458 & ch278             & $-22.28 \pm 0.05$         & $42.79 \pm 1.39$          & $324.51 \pm 1.91$         & A & 4.9 & 187.30  & 1.51  \\ 
G351.584$-$00.350 & ch279             & $-92.65 \pm 0.11$         & $5.74 \pm 1.85$           & $101.29 \pm 14.09$        & A & 2.9 & 100.10  & 0.25  \\ 
G351.620+00.143   & ch280             & $-44.40 \pm 0.05$         & $38.61 \pm 1.06$          & $60.31 \pm 1.52$          & A & 5.6 & 342.60  & 1.66  \\ 
G351.646$-$01.252 & ch282             & $-13.71 \pm 0.06$         & $25.43 \pm 1.13$          & $34.91 \pm 2.45$          & A & 4.8 & 498.30  & 1.79  \\ 
G351.688$-$01.169 & ch283             & $-12.55 \pm 0.04$         & $52.30 \pm 0.74$          & $324.71 \pm 0.77$         & B & 6.6 & 200.10  & 1.64  \\ 
G352.597$-$00.188 & ch286             & $-80.75 \pm 0.07$         & $12.68 \pm 0.81$          & $3.09 \pm 4.73$           & A & 3.4 & 136.10  & 0.33  \\ 
G353.408$-$00.381 & ch293             & $-15.83 \pm 0.05$         & $4.53 \pm 0.65$           & $120.48 \pm 7.61$         & A & 3.9 & 253.80  & 0.13  \\ 
G354.175$-$00.062 & ch297             & $-32.69 \pm 0.08$         & $16.95 \pm 1.10$          & $127.29 \pm 3.39$         & A & 3.7 & 125.90  & 0.52  \\ 
G354.465+00.079   & ch298             & $18.69 \pm 0.12$          & $15.97 \pm 1.69$          & $114.86 \pm 4.93$         & B & 3.6 & 93.50   & 1.57  \\ 
G354.936+00.330   & ch301             & $17.88 \pm 0.27$          & $15.85 \pm 4.87$          & $230.81 \pm 16.27$        & A & 1.9 & 42.60   & 1.16  \\ 
G356.230+00.670   & ch303             & $116.85 \pm 0.06$         & $21.25 \pm 0.91$          & $47.12 \pm 2.47$          & A & 4.2 & 139.70  & 0.55  \\ 
G356.310$-$00.206 & ch304             & $-5.82 \pm 0.14$          & $8.21 \pm 2.35$           & $163.85 \pm 16.15$        & A & 3.0 & 75.70   & 0.39  \\ 
G358.633+00.062   & fa054             & $13.40 \pm 0.33$          & $70.13 \pm 13.35$         & $144.48 \pm 10.89$        & A & 1.6 & 29.10   & 1.79  \\ 
  \enddata 
  \tablecomments{MLE velocity gradient fit includes the velocity 
    offset, $c_{0}$ in \kms, the velocity gradient magnitude, 
    $\nabla V_{\rm LSR}$ in \msarcsec, and the position angle, PA in 
    degrees. QF is the quality factor (see text) and RMSE is the root mean square error in \kms.} 
\end{deluxetable}

\newpage

\begin{deluxetable}{llrcccccc} 
  \tabletypesize{\scriptsize} 
  \tablecaption{VLA \hii\ Region Sample}\label{tab:vla} 
  \tablewidth{0pt} 
  \setlength{\tabcolsep}{2.0pt} 
  \tablehead{\colhead{} & \colhead{} & 
    \multicolumn{3}{c}{\underline{~~~~~~~~~~~MLE Velocity Gradient Fit~~~~~~~~~~~}} & 
    \colhead{} & \colhead{} & \colhead{} & \colhead{} \\ 
    \colhead{} & \colhead{} & 
    \colhead{$c_{0}$} & \colhead{$\nabla V_{\rm LSR}$} & \colhead{PA} & 
    \colhead{} & \colhead{No.} & \colhead{Peak} & \colhead{RMSE} \\ 
    \colhead{Source} & \colhead{Field} & 
    \colhead{(\kms)} & \colhead{(\msarcsec)} & \colhead{(degree)} & 
    \colhead{QF} & \colhead{Beams} & \colhead{SNR} & \colhead{(\kms)} } 
  \startdata 
G013.880+00.285   & G013.880+0.285    & $49.24 \pm 0.02$          & $111.61 \pm 1.03$         & $201.89 \pm 0.82$         & A & 9.6 & 415.70  & 1.06  \\ 
G027.562+00.084   & G027.562+0.084    & $87.93 \pm 0.28$          & $149.75 \pm 32.19$        & $277.05 \pm 12.55$        & A & 1.4 & 29.20   & 0.25  \\ 
G034.133+00.471   & G034.133+0.471    & $35.67 \pm 0.11$          & $166.13 \pm 13.98$        & $104.67 \pm 5.18$         & A & 2.8 & 95.50   & 0.63  \\ 
G035.126$-$00.755 & G035.126$-$0.755  & $42.69 \pm 0.15$          & $166.97 \pm 3.36$         & $270.37 \pm 1.81$         & A & 2.8 & 44.00   & 0.38  \\ 
G038.550+00.163   & G038.550+0.163    & $27.76 \pm 0.30$          & $205.27 \pm 44.18$        & $156.12 \pm 12.07$        & A & 1.2 & 31.20   & 0.18  \\ 
G049.399$-$00.490 & G049.399$-$0.490  & $61.64 \pm 0.13$          & $82.72 \pm 11.26$         & $48.33 \pm 7.86$          & A & 3.8 & 48.90   & 0.77  \\ 
G070.280+01.583   & G070.280+1.583    & $-25.05 \pm 0.09$         & $165.19 \pm 6.02$         & $259.92 \pm 1.90$         & A & 7.1 & 56.50   & 1.05  \\ 
G070.329+01.589   & G070.280+1.583    & $-25.02 \pm 0.19$         & $110.08 \pm 1.86$         & $314.30 \pm 0.96$         & B & 2.3 & 61.80   & 0.89  \\ 
G075.768+00.344   & G075.768+0.344    & $-5.19 \pm 0.04$          & $309.53 \pm 2.73$         & $347.90 \pm 0.58$         & A & 11.2 & 154.10  & 1.70  \\ 
G097.515+03.173   & G097.515+3.173    & $-75.88 \pm 0.21$         & $667.10 \pm 45.61$        & $244.42 \pm 2.38$         & A & 3.7 & 32.40   & 6.07  \\ 
G351.246+00.673   & G351.246+0.673    & $0.91 \pm 0.03$           & $98.99 \pm 1.31$          & $241.10 \pm 0.71$         & A & 9.0 & 313.00  & 2.46  \\ 
G351.311+00.663   & G351.311+0.663    & $-5.16 \pm 0.04$          & $124.64 \pm 1.49$         & $284.61 \pm 0.67$         & A & 6.4 & 195.00  & 0.94  \\ 
  \enddata 
  \tablecomments{MLE velocity gradient fit includes the velocity 
    offset, $c_{0}$ in \kms, the velocity gradient magnitude, 
    $\nabla V_{\rm LSR}$ in \msarcsec, and the position angle, PA in 
    degrees. QF is the quality factor (see text) and RMSE is the root mean square error in \kms.} 
\end{deluxetable}

There is no correlation between the velocity gradient magnitude and
position angle.  We have converted the position angles from Equatorial
to Galactic coordinates to test whether there is a physical connection
between Galactic rotation and a preferred rotation axis of these
nebulae.  The position angles appear to have a random distribution on
the sky with no preferred direction with respect to the Galactic Plane
(see Figure~\ref{fig:grad-fits}).

\begin{figure}
  \centering
  \includegraphics[angle=0,scale=0.8]{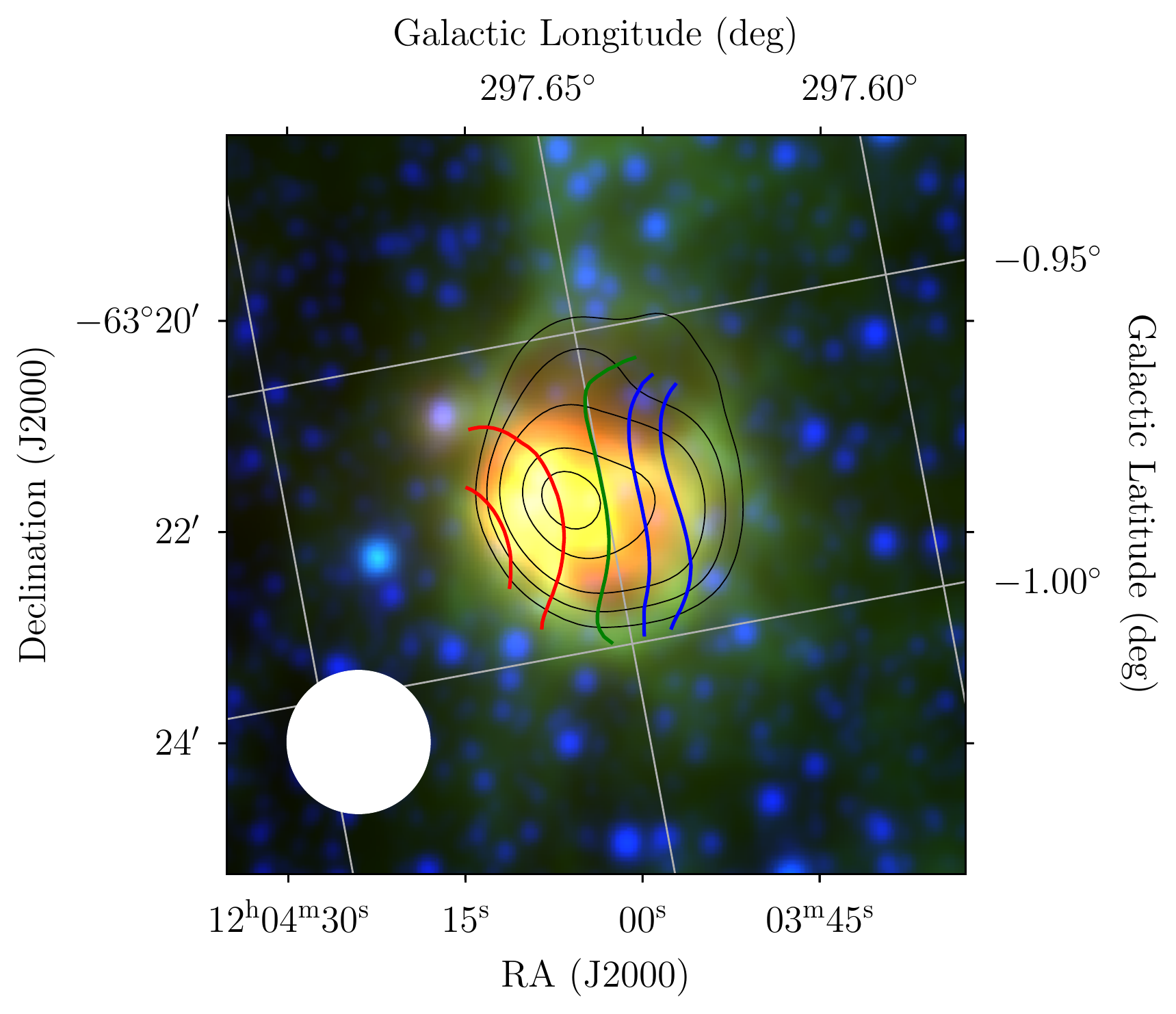}
  \caption{WISE mid-infrared image of G297.651$-$00.973.  The
    photodissociation region (PDR) is visible at 12\micron\ (green)
    from PAH emission and surrounds the \hii\ region detected at
    22\micron\ (red) from heated dust grains within the ionized gas.
    The 3.4\micron\ (blue) emission is from low-mass stars.  The ATCA
    free-free radio continuum emission is overlaid on the mid-infrared
    image as black contours at 5\%, 10\%, 20\%, 50\%, and 80\% of the
    peak continuum brightness.  The \hna\ RRL $V_{\rm LSR}$ velocity
    gradient structure is shown by the colored contours.  The green
    contour corresponds to the Gaussian center $V_{\rm LSR}$ fitted to
    the peak spectrum. The redshifted and blueshifted contours are
    spaced equally between the fitted peak Gaussian $V_{\rm LSR}$ and
    the maximum redshift/blueshift.  The HPBW size is shown by the
    white circle in the bottom left-hand corner of the image.}
  \label{fig:mle-atca}
\end{figure}

\begin{figure}
  \centering
  \includegraphics[angle=0,scale=0.8]{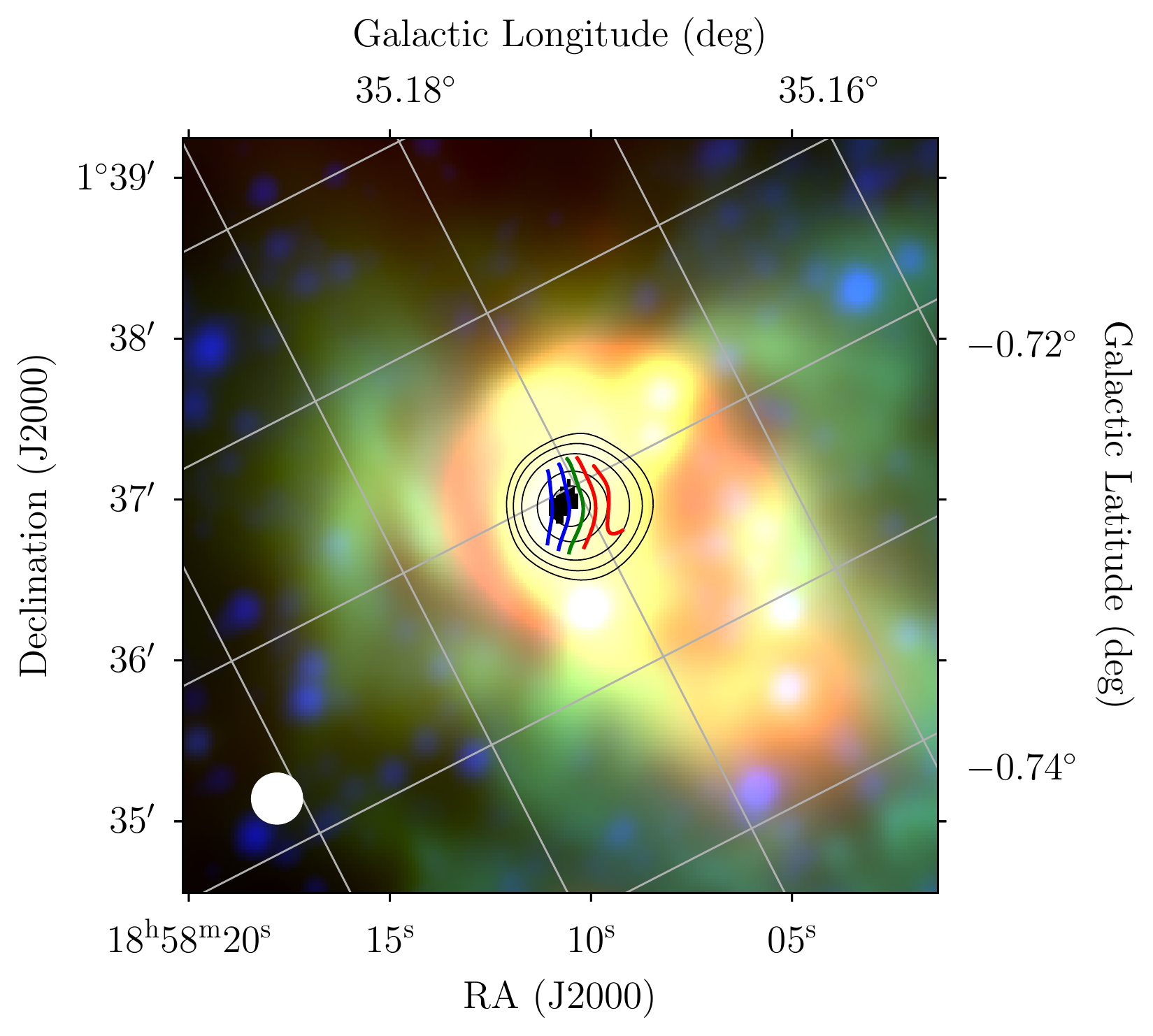}
  \caption{WISE mid-infrared image of G035.126$-$0.755.  See
    Figure~\ref{fig:mle-atca} for details except the radio data are
    from the VLA.}
\label{fig:mle-vla}
\end{figure}

We produce composite mid-infrared/radio images for each source to help
interpret the kinematic structure.  We combine data from the
Wide-Field Infrared Survey Explorer (WISE) with RRL and continuum data
from either the ATCA or VLA.  Examples are shown in
Figures~\ref{fig:mle-atca} and \ref{fig:mle-vla} for the ATCA \hii\
region G297.651$-$00.973 and the VLA \hii\ region G035.126$-$0.755,
respectively.  The color image corresponds to the WISE mid-infrared
emission.  \hii\ regions, detected in 22\micron\ (red) emission from
heated dust grains, are often surrounded by a photodissociation
region, visible in 12\micron\ (green) emission from polycyclic
aromatic hydrocarbon (PAH) molecules.  Overlaid on the WISE images are
black contours from the free-free radio continuum that shows the
extent of the \hii\ region.  The \hna\ RRL Gaussian fit center
velocity $V_{\rm LSR}$ is represented with color contours and
indicates the direction of the velocity gradient.

Some sources, such as G297.651$-$00.973 (Figure~\ref{fig:mle-atca}),
show a bipolar morphology in the infrared emission.  If the kinematics
detected by the RRL emission were due to ionized gas motions within
the bipolar outflow, we would expect the velocity gradient to be
aligned with the bipolar outflow axis.  For G297.651$-$00.973, the
velocity gradient is perpendicular to the outflow direction, however,
consistent with ionized gas rotating around the bipolar outflow; that
is, the axis of rotation is along the direction of the bipolar
outflow.

Another common infrared morphology is bubble-like structures.  For
G035.126$-$0.755 (Figure~\ref{fig:mle-vla}), we see a complex bubble
morphology surrounding the radio continuum emission.  We do not have
the spatial resolution to infer any connection between the RRL
velocity gradient and the bubble structure, but bubble infrared
morphologies are common toward \hii\ regions
\citep[e.g.,][]{anderson11}.

Since bulk motions can modify the line width we also inspect the
morphology of the $\Delta{V}$ images.  For a static nebula we expect
the line width to be uniform, but we detect structure in $\Delta{V}$
images in about half of our nebulae.  For example, in both
G297.651$-$00.973 and G035.126$-$0.755 the line widths are higher near
the center of the \hii\ region; that is, centrally peaked.  The
combination of a velocity gradient with a centrally peaked line width
was also detected in the \hii\ region K3-50A \citep{balser01}.  We
detect centrally peaked line widths in about 15\% of the sources in
Tables~\ref{tab:atca} and \ref{tab:vla}.  About 15\% of nebulae have
higher line widths toward one side of the \hii\ region boundary, or
edge peaked.  We also see a gradient in $\Delta{V}$ across nebulae in
about 20\% of the sources in our sample.  The RRL line width
morphology of the remaining sources is roughly split between uniform
and complex.

\clearpage

\section{\hii\ Region Simulations}\label{sec:sims}

To better interpret these results, we develop \hii\ region numerical
simulations that include a dynamical model to explain the observed
kinematics.  Assuming an optically thin RRL in Local Thermodynamical
Equilibrium (LTE), what type of dynamical model could produce RRL
velocity gradients?  The thermal gradient at the boundary of a
spherical, homogeneous \hii\ region would produce a spherically
symmetric expanding nebula \citep[e.g.,][]{spitzer78, dyson97}.
Radiation and winds from OB-type stars could also cause expansion in
young, compact \hii\ regions \citep[e.g.,][]{oey96, lopez14, rugel19,
  mcleod19}.  But a spherically symmetric expanding nebula would
produce double-peaked RRL profiles toward the center of the nebula,
or, if the nebula were unresolved, a square-wave profile
\citep[e.g.,][]{balser97}.  We do not detect any such profiles in the
ATCA or VLA RRL data cubes, but we do see bipolar outflow morphologies
in some of the mid-infrared images.

Here, we explore bipolar outflows and solid body rotation, although
there may be other \hii\ region dynamics that can produce velocity
gradients.  Moreover, complex velocity structures with small spatial
scales could produce velocity gradients when convolved with the
telescope's larger beam.  An ionized bipolar outflow could produce a
RRL velocity gradient with the velocity gradient direction aligned
with the outflow axis.  This depends on the relative orientation of
the bipolar outflow axis with respect to the observer.  A rotating
spherical nebula could also produce a RRL velocity gradient where the
velocity gradient direction would be normal to the rotation axis.

We assume a homogeneous, isothermal, spherical nebula and perform the
radiative transfer of RRL and continuum emission to produce a model
brightness temperature distribution on the sky.  The spectral noise is
modeled as random (Gaussian) noise with a specified RMS.  We then
generate synthetic RRL spectra by convolving the brightness
temperature with a telescope beam.  The details are given in
Appendix~\ref{sec:model}.  These simulations produce a RRL
position-position-velocity data cube. We analyze these synthetic RRL
data cubes using the same procedures as for our ATCA and VLA data
discussed in Section~\ref{sec:kinematics}.  We fit a single Gaussian
component within each spaxel of the data cube and generate images of
the center velocity, $V_{\rm LSR}$, and FWHM line width, $\Delta V$.

To investigate the effects of either bipolar outflows or solid body
rotational motions, we run a series of simulations exploring a range
of \hii\ region kinematic model parameters (see Table~\ref{tab:sim}).
The outflow and rotation speeds are somewhat arbitrary. Nevertheless,
hydrodynamic simulations produce outflow speeds \citep{bodenheimer79}
and rotational speeds \citep{jeffreson20} within the range of values
explored in Table~\ref{tab:sim}.  We only consider nebulae with a
diameter of 2\pc\ at a distance of either 2\kpc\ or 5\kpc, producing
angular \hii\ region sizes of 206\arcsec\ or 83\arcsec, respectively.
Since our synthetic telescope HPBW is 90\arcsec, this corresponds to a
slightly resolved or unresolved source, similar to the conditions for
many of the sources in our \hii\ region sample.

\begin{deluxetable}{lcl}
\tablecaption{\hii\ Region Kinematic Model Parameter Ranges\label{tab:sim}}
\tablehead{
\colhead{Parameter} & \colhead{Range} & \colhead{Comment}
}
\startdata
$V_{\rm o}$ & 20--120\kms\ & Outflow speed. \\
$\Theta_{\rm o}$ & 15--85\degree\ & Outflow opening angle. \\
$V_{\rm eq}$ & 5--100\kms\ & Rotation equatorial speed. \\
$\Phi_{\rm los}$ & 0--90\degree\ & Outflow/Rotation axis relative to the line-of-sight. \\
$\Phi_{\rm sky}$ & 0--90\degree\ & Position angle of the Outflow/Rotation axis on the sky. \\
\enddata
%\tablecomments{} 
%\tablenotetext{a}{}
\end{deluxetable}

\begin{figure}
  \centering
  \includegraphics[angle=0,scale=0.60]{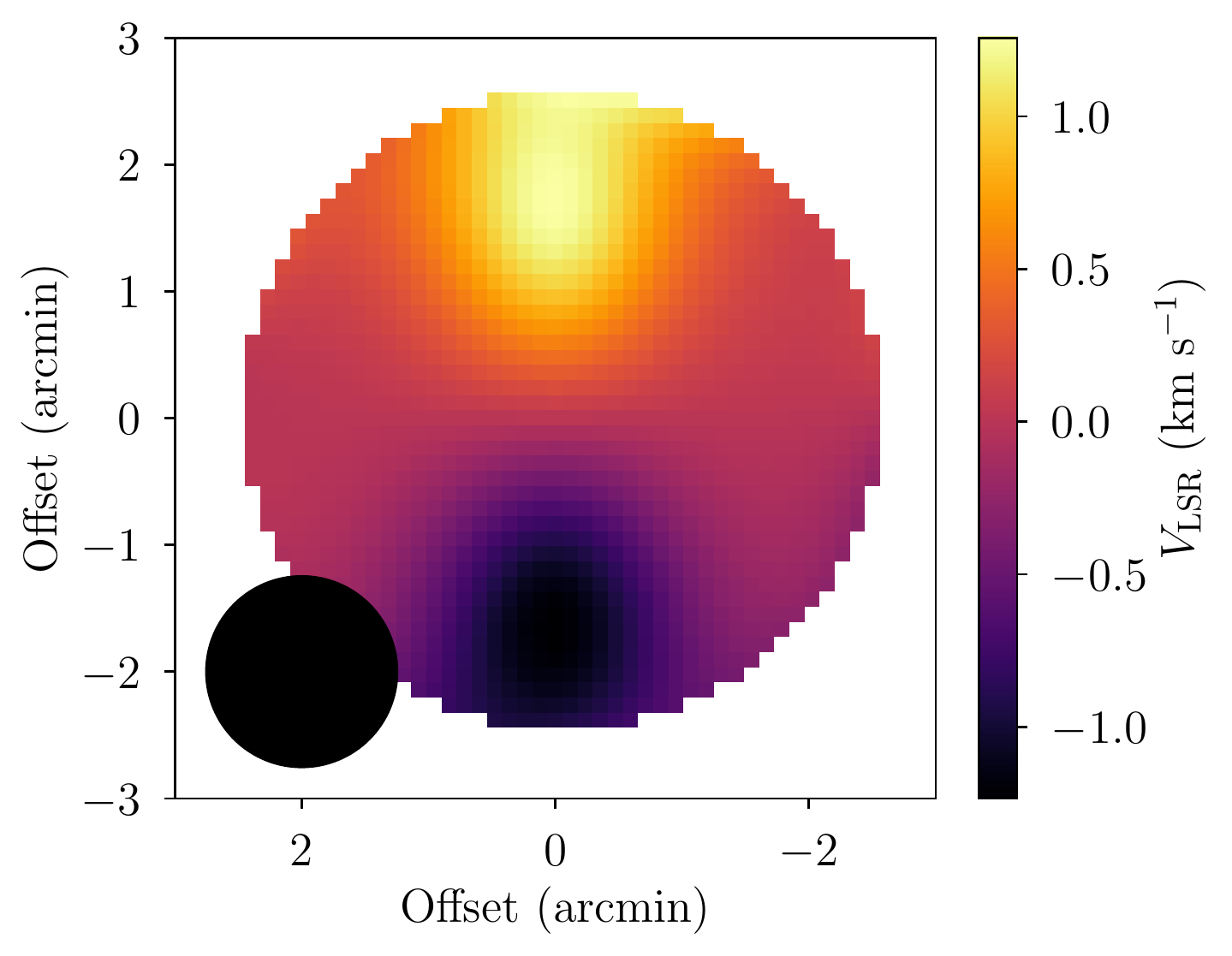}
  \includegraphics[angle=0,scale=0.60]{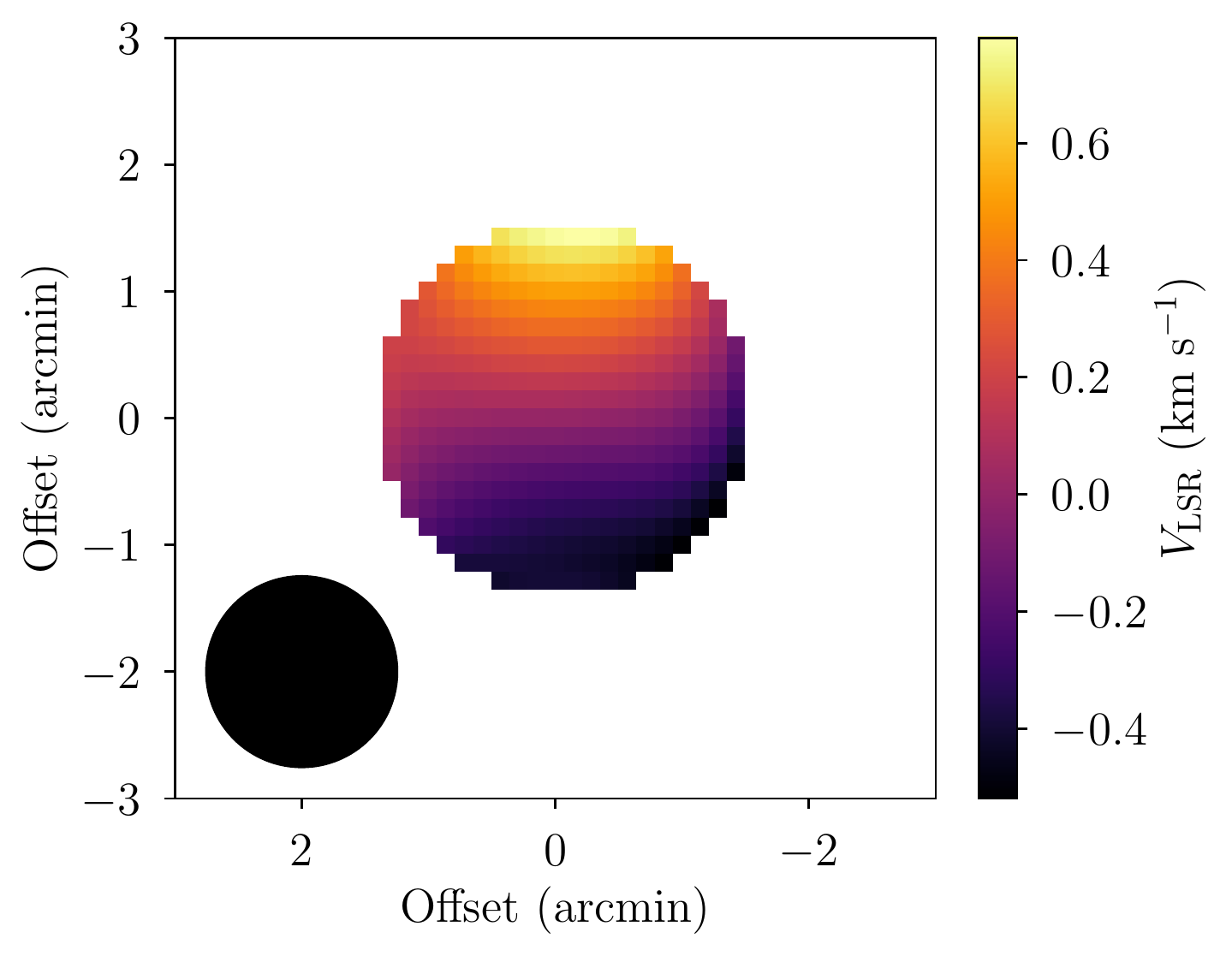}
  \includegraphics[angle=0,scale=0.60]{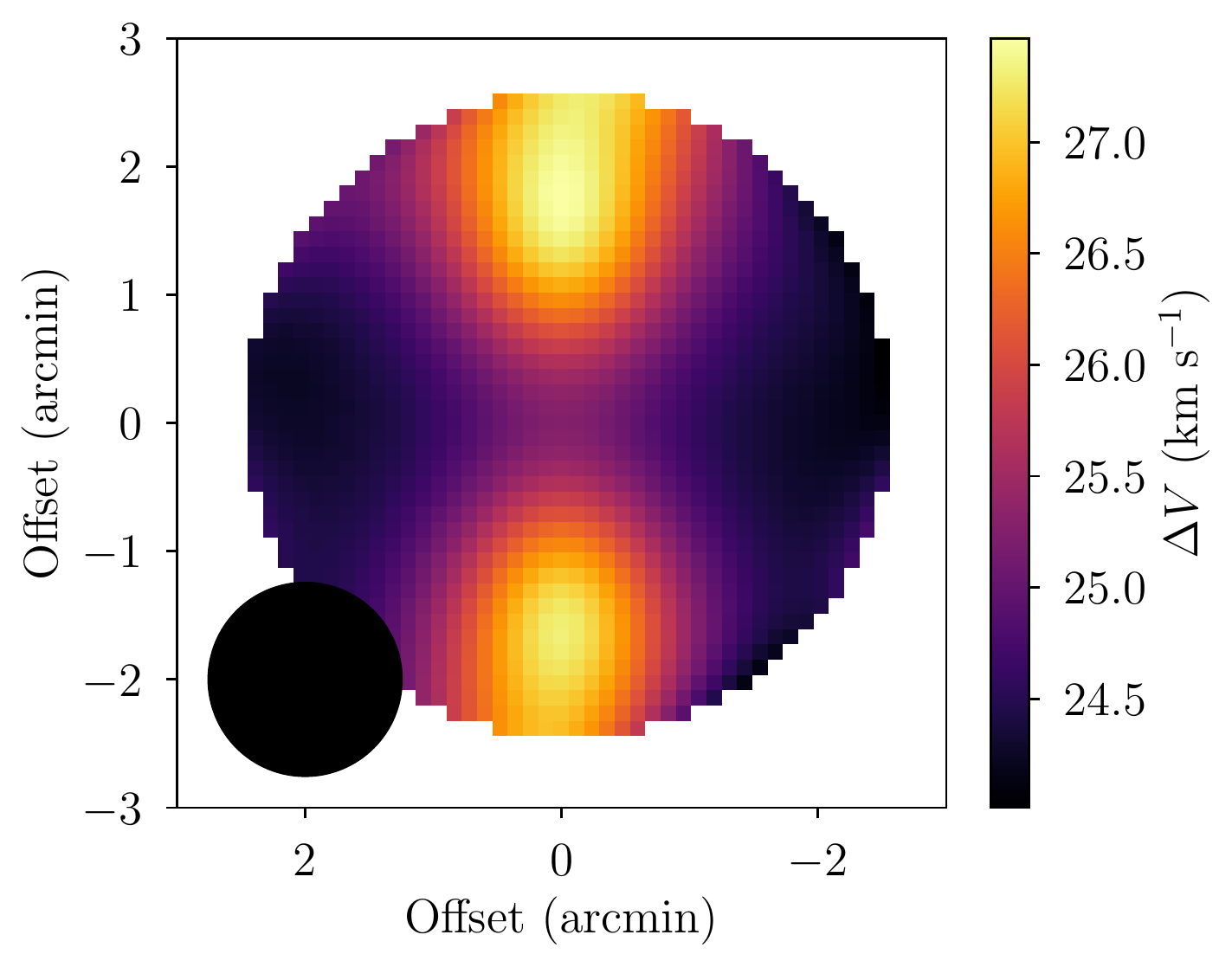}
  \includegraphics[angle=0,scale=0.60]{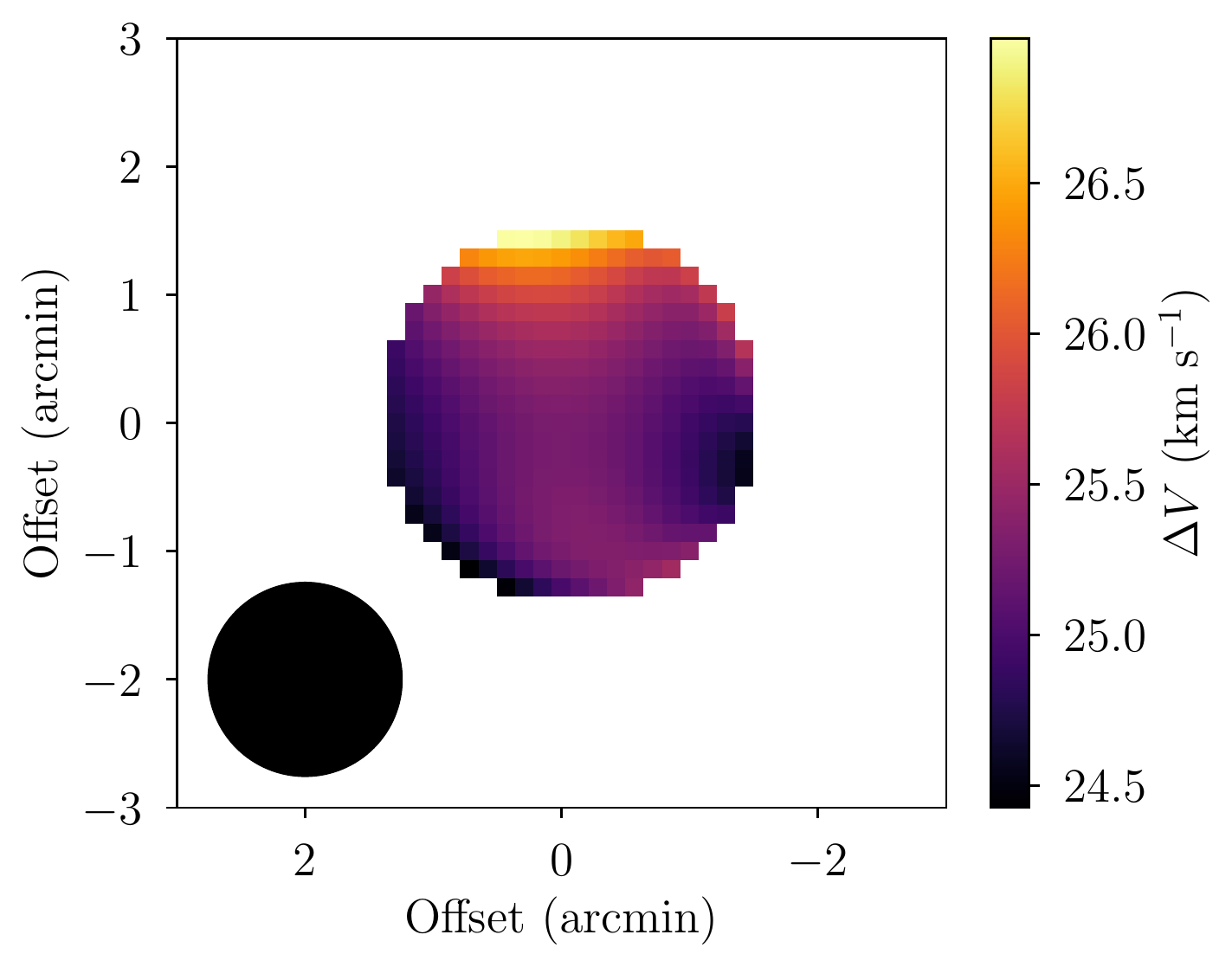}
  \caption{Bipolar outflow simulation results.  The center velocity,
    $V_{\rm LSR}$, (top) and FWHM line width, $\Delta{V}$, (bottom)
    images are shown for two simulations with identical model nebulae
    at different distances.  The {\it resolved} model nebula at a
    distance of 2\kpc\ has an angular size of 206\arcsec\ (left),
    whereas the {\it unresolved} model nebula at a distance of 5\kpc\
    has an angular size of 83\arcsec\ (right).  The model telescope
    HPBW of 90\arcsec\ is shown in the bottom left-hand corner of the
    image.  The bipolar outflow model assumes an outflow velocity
    $V_{\rm o} = 40$\kms\ with opening angle of
    $\Theta_{\rm o} = 45$\degree.  The bipolar outflow axis is tilted
    45\degree\ toward the observer, $\Phi_{\rm los} = 45$\degree, and
    oriented north on the sky, $\Phi_{\rm sky} = 0$\degree.}
\label{fig:outflow}
\end{figure}

\begin{figure}
  \centering
  \includegraphics[angle=0,scale=0.60]{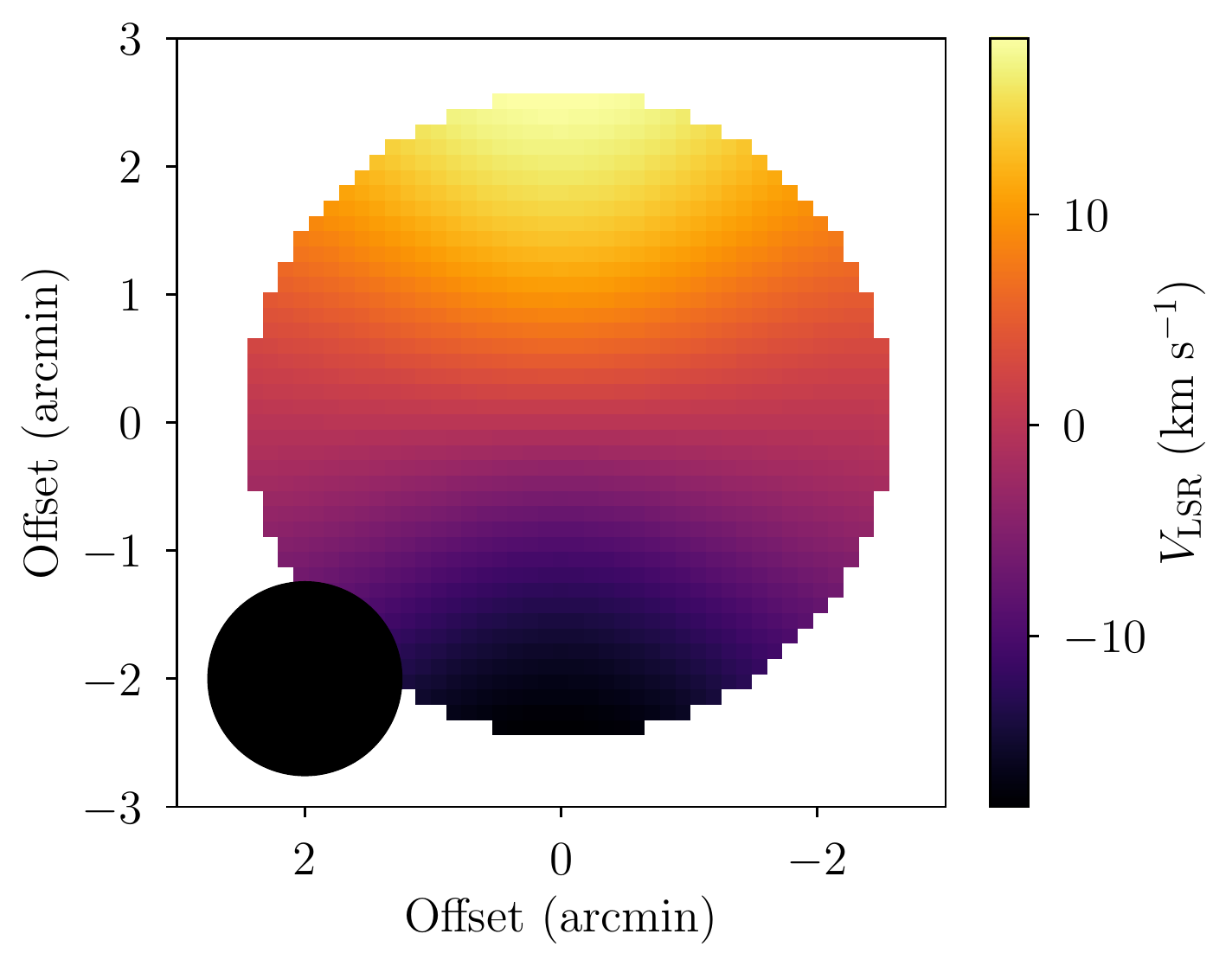}
  \includegraphics[angle=0,scale=0.60]{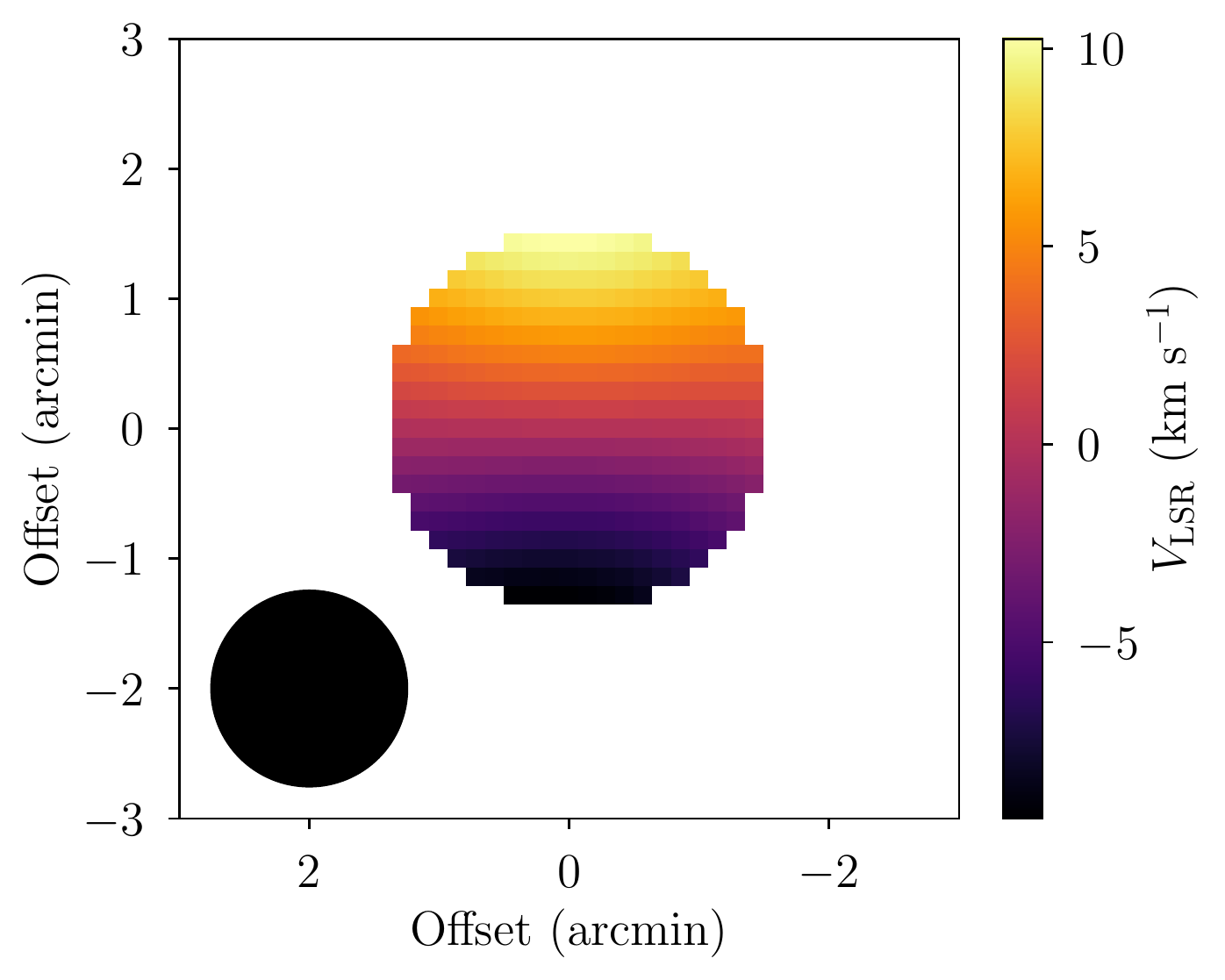}
  \includegraphics[angle=0,scale=0.60]{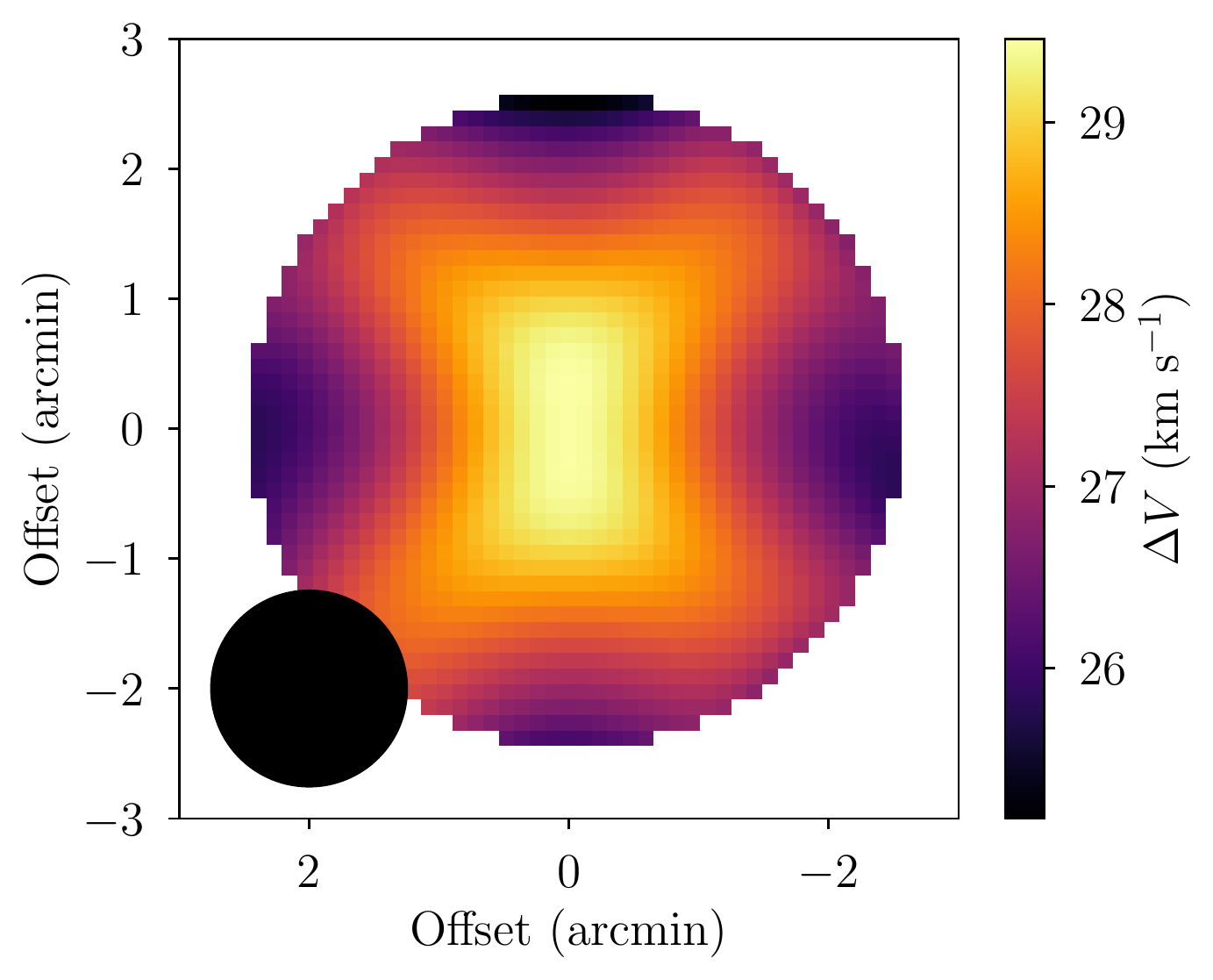}
  \includegraphics[angle=0,scale=0.60]{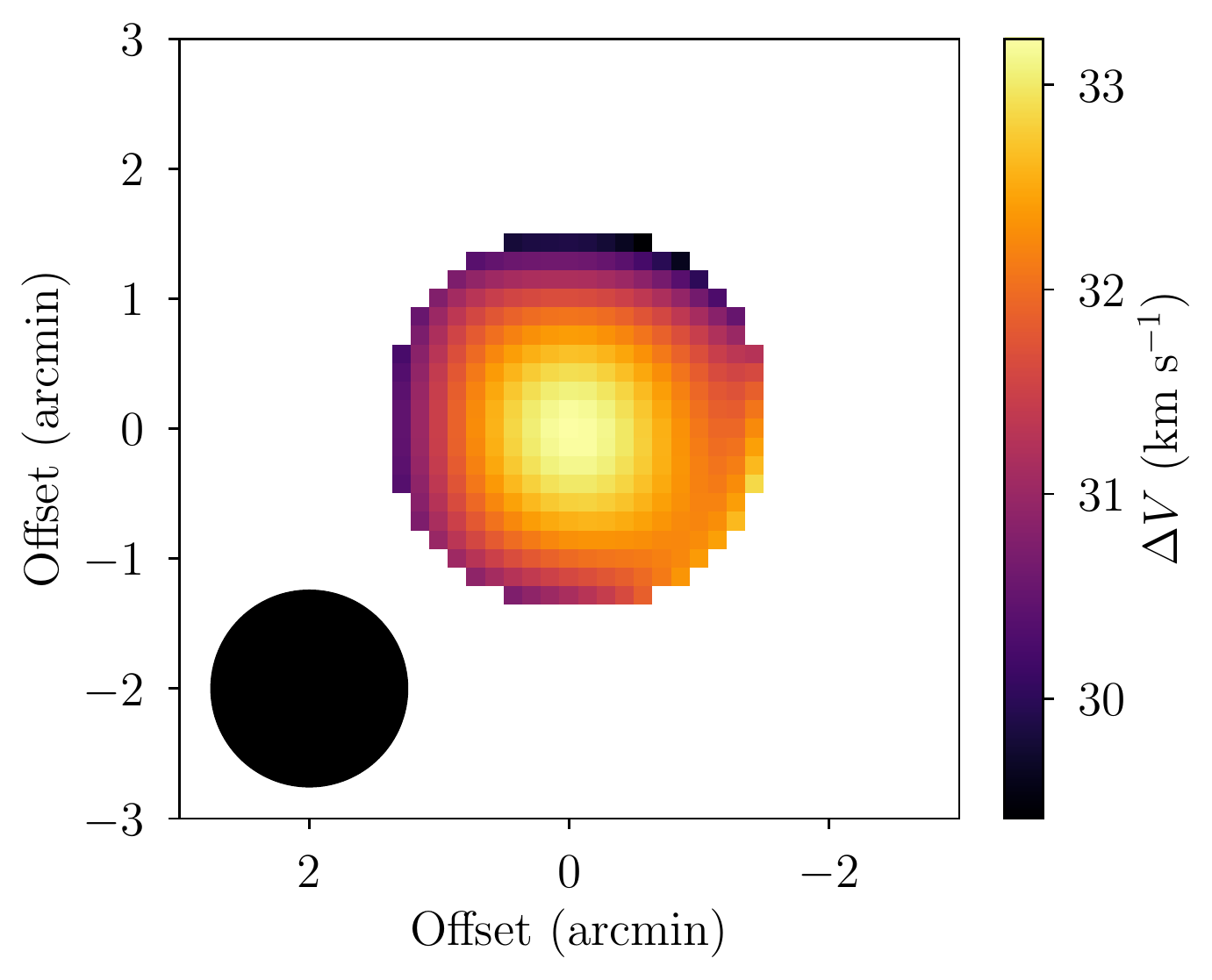}
  \caption{Solid body rotation simulation results.  The center
    velocity, $V_{\rm LSR}$, (top) and FWHM line width, $\Delta{V}$,
    (bottom) images are shown for two simulations with identical model
    nebula at different distances.  The {\it resolved} model nebulae
    at a distance of 2\kpc\ has an angular size of 206\arcsec\ (left),
    whereas the {\it unresolved} model nebula at a distance of 5\kpc\
    has an angular size of 83\arcsec\ (right).  The model telescope
    HPBW of 90\arcsec\ is shown in the bottom left-hand corner of the
    image.  The solid body rotation model assumes an equatorial speed
    of $V_{\rm eq} = 45$\kms.  The rotation axis is perpendicular to
    the observer, $\Phi_{\rm los} = 0$\degree, and oriented East on
    the sky, $\Phi_{\rm sky} = 90$\degree.}
\label{fig:solidbody}
\end{figure}

The main results are illustrated in Figures~\ref{fig:outflow} and
\ref{fig:solidbody} for the bipolar outflow and solid body rotation
models, respectively.  For each dynamical model we show the results of
a simulation where the \hii\ region is resolved (left panels) and a
simulation where the \hii\ region is slightly unresolved (right
panels); that is, the nebular angular size is less than the HPBW.
Velocity gradients are detected for both the bipolar outflow and solid
body rotation model simulations.  When the nebula is unresolved the
synthetic $V_{\rm LSR}$ images for the bipolar outflow and solid body
rotation models are very similar.  The source emission is larger than
the synthetic HPBW since we are detecting emission at the edge of the
beam.  The velocity gradient magnitude increases with better spatial
resolution.

There are several notable differences between the two dynamical
models.  When the nebula is resolved the bipolar morphology can be
seen in both the $V_{\rm LSR}$ and $\Delta V$ images, given sufficient
sensitivity.  The solid body rotation models have a centrally peaked
line width because along the center direction there is a high gradient
in velocity which is blended by the beam into a broad line width.

The solid body rotation models can produce larger velocity gradients
than the bipolar outflow models.  This is because there are two main
factors that limit the velocity gradient magnitude in the bipolar
outflow models.
\begin{enumerate}

\item {\it Projection Effect.} There is always a significant
  projection effect of the outflow axis with respect to the
  line-of-sight.  Velocity gradients are not visible if the outflow
  axis is pointed toward the observer ($\Phi_{\rm los} = 90$\degree)
  due to symmetry, or pointed normal to the line-of-sight
  ($\Phi_{\rm los} = 0$\degree) since there is no radial velocity
  component.  The optimal orientation to detect a velocity gradient is
  when $\Phi_{\rm los} = 45$\degree, where the outflow is observed in
  projection.  In contrast, there are no projection effects for the
  solid body rotation model when the rotation axis is normal to the
  line-of-sight ($\Phi_{\rm los} = 0$\degree).  The velocity gradient
  does decrease as $\Phi_{\rm los}$ increases for the solid body
  rotation model, but there are orientations with small (or zero)
  projection effects.

\item {\it Filling Factor Effect.} Since the bipolar outflow model
  does not fill the area of the nebula as projected onto the sky, the
  motions of the bipolar outflow are diluted; that is, there is a
  filling factor that is less than one.  Increasing the bipolar
  outflow opening angle mitigates this effect to some degree, but the
  magnitude of the measured velocity gradient is constrained by the
  RRL width.  Most of the nebular emission resides at the systemic RRL
  velocity, so as the outflow speed is increased the emission arsing
  from the bipolar outflow region moves to the wings of the main
  Gaussian profile.  This is not true for the solid body rotation
  models since all of the gas is rotating.

\end{enumerate}

The simulations explain some of the results from our ATCA and VLA
observations that are discussed in Section~\ref{sec:kinematics}.  We
detect velocity gradients even when the source is not resolved because
our RRL data have enough sensitivity to detect emission at the edge of
the beam.  We observe two sources with both the ATCA and VLA:
G013.880+00.285 and G351.246+00.673 (see
Tables~\ref{tab:atca}--\ref{tab:vla}).  The velocity gradient
magnitude increases with better spatial resolution, consistent with
the simulations.  A caveat is that the ATCA and VLA sample different
spatial scales and therefore different zones of the nebula.  We detect
centrally peaked RRL widths in about 15\% of the sources in our
sample, consistent with the solid body rotation model.

Overall, the simulations favor the solid body rotation model over the
bipolar outflow model for several reasons.  First, we observe velocity
gradients significantly larger than expected for the bipolar outflow
model.  The simulations are restricted to model nebulae that are
either slightly resolved or unresolved, similar to our ATCA \hii\
region sample in Table~\ref{tab:atca} where the angular sizes are
similar to the HPBW.  (If the source size were significantly smaller
than the HPBW, we would not be able to detect any velocity structure.)
The maximum velocity gradient produced by our bipolar outflow model
simulations is about 40\msarcsec, yet we detect velocity gradients
significantly higher in the ATCA \hii\ region sample.  The parameters
for the bipolar outflow model with the maximum velocity gradient are:
$V_{\rm o} = 30$\kms, $\Theta_{\rm o} = 85$\degree, and
$\Phi_{\rm los} = 45$\degree.  For this model the nebula is
unresolved.  Increasing the outflow velocity beyond
$V_{\rm o} = 30$\kms\ reduces the velocity gradient because of the
filling factor effect discussed above.  In contrast, there is no such
limitation for the solid body rotation model because all of the gas is
rotating.  For solid body rotation with $V_{\rm eq} = 15$\kms\ and
$\Phi_{\rm los} = 0$\degree, the simulation produces a velocity
gradient of $\nabla V_{\rm LSR} = 83$\msarcsec\ for an unresolved
nebula.  Increasing the equatorial rotation speed increases the
velocity gradient.  Second, the bipolar outflow model simulations
predict that we should see a bipolar morphology when the nebula is
resolved, but we do not observe any bipolar structure in the
$V_{\rm LSR}$ images.  Lastly, we detect a centrally peaked RRL width
in about 15\% of our sources, consistent with the solid body rotation
model.

The model \hii\ region simulations discussed here are limited.  They
only include simple dynamical models for bipolar outflows and
rotational motions.  Moreover, we have not explored the full model
parameter phase space. More sophisticated Monte Carlo simulations
could make predictions about the expected distribution of velocity
gradient magnitudes, but this is beyond the scope of this paper.
Nevertheless, the solid body rotation models are better able to
reproduce the velocity gradient properties determined from our \hii\
regions RRL observations.

\section{Discussion}\label{sec:discussion}

Rotation is a common property of most objects in the Milky Way
\citep{belloche13}.  On the largest scales there is differential
rotation in the Galactic disk.  Simulations indicate that the
rotational properties of \hi\ clouds and GMCs are correlated with the
time scale for Galactic shearing and the gravitational free-fall time
\citep{jeffreson20}.  The specific angular momentum of the stars and
planets that form within GMCs is about six orders of magnitude less
than the dense cores in GMCs \citep{belloche13}.  Since angular
momentum is conserved there must be a transfer of angular moment
during star formation, but the possible candidates (e.g.,
fragmentation, magnetic braking, etc.)  cannot explain the difference;
hence the ``Angular Momentum Problem'' in star formation
\citep{spitzer78, bodenheimer95, mathieu04}.

Where do the dense cores in GMCs obtain their angular momentum?
Results from Herschel indicate that most molecular cores form in
filaments with a typical width of 0.1\pc\ \citep{andre10, andre14,
  koch15}.  These molecular filaments tend to have sizes between the
GMCs and the molecular cores.  \citet{hsieh21} performed observations
of N$_{2}$D$^{+}$ in Orion B and detect a velocity gradient along the
molecular filament's minor axis.  The authors argue that the velocity
gradient is due to rotation of the filament and that the derived
angular momentum profile, how the angular momentum varies with
distance from the center of the filament, is consistent with ambient
turbulence as the source of the filament's rotation
\citep[see][]{misugi19}. Tracing the angular momentum through the
stages of gravitational collapse that lead to high mass star formation
is a fundamental problem that calls for further simulations and
observations with various ISM tracers.

\citet{dalgleish18} use the simulations of \citet{peters10a} to
develop a cartoon picture of how \hii\ regions kinematically evolve.
\citet{peters10a} use the FLASH code to develop three-dimensional
simulations of a rotating, collapsing molecular cloud and include
heating from both ionizing and non-ionizing sources \citep[also
see][]{peters10b, peters10c}.  The simulation begins with a molecular
cloud which has a net angular momentum.  Stars form in the denser
regions and follow the rotation of the cloud.  The angular momentum of
the \hii\ region is inherited from the parent cloud only indirectly,
via these massive ionizing stars, whose motions within the expanding
\hii\ region complicate the dynamics of the ionized gas.
Nevertheless, \citet{dalgleish18} predict that most \hii\ regions
should contain the kinematic signature of rotation---velocity
gradients.

We detect velocity gradients between
$\nabla V_{\rm LSR} = 5-200$\msarcsec\ in about 49\% of the sources in
our \hii\ region sample.  Moreover, the RRL widths are centrally
peaked in about 15\% of the sources with velocity gradients.  Simple
models of solid body rotation can produce velocity gradients with
similar magnitudes as observed and have centrally peaked line widths.
These results are all consistent with the hypothesis that most \hii\
regions have inherited some of the angular momentum from their
rotating parent molecular clouds.

We cannot, however, rule out other dynamical models.  We have shown
that nebula models of bipolar outflow also predict velocity gradients.
Moreover, bipolar outflow structure is observed in infrared emission
toward some sources in our sample with detected velocity gradients.
If the velocity gradient were due to the bipolar outflow motions then
the velocity gradient should be along the direction of the bipolar
outflow axis.  Another possibility is that the velocity gradient is
primarily due to rotation from a disk of material and the bipolar
outflow is aligned with the rotation axis (e.g., G297.651$-$00.973 in
Figure~\ref{fig:mle-atca}).  We only detect an infrared bipolar
morphology in about 10 sources from our sample and there is no
consistent alignment between the bipolar morphology axis and the RRL
velocity gradient direction.  There are examples in the literature
where there is evidence for both rotation and a bipolar outflow in the
ionized gas: \ngc{6334A} \citep{depree95, balser01}, K3-50A
\citep{depree94, balser01}, W49A \citep{mufson77, welch87}, and
G316.81–0.06 \citep{dalgleish18}.

To distinguish between these two dynamical models requires better
spatial resolution.  In some cases both rotation and outflow may be
present but they may occur on different spatial scales.  The \hii\
region \ngc{6334A} is a good example where the bipolar outflow extends
beyond the core region and the kinematics can be separated with
position-velocity maps \citep{balser01}.  Regardless, the simple
bipolar outflow model developed here cannot explain many of the
velocity gradients in our sample because these models are unable to
produce the large velocity gradients or the centrally peaked line
widths.  More complex models could overcome some of these limitations.
For example, a bipolar outflow \hii\ region where the density is
significantly enhanced in the outflow region.

\citet{dalgleish18} speculate that if Galactic rotation provides the
initial angular momentum of molecular clouds then there should be a
connection between the orientation of the angular momentum axis of the
molecular cloud and the Galactic plane \citep[also see][]{garay86}.
Figure~\ref{fig:grad-fits} plots the velocity gradient position angle
distribution with respect to Galactic Plane.  Since the position angle
distribution is approximately random, there is no evidence that the
initial angular momentum of the molecular clouds from which the \hii\
regions formed is set by Galactic rotation.

The ATCA and VLA RRL surveys provide a nice sample to explore \hii\
region kinematics and provide constraints on how angular momentum
governs star formation.  More sensitive, higher spatial resolution
images of the \hii\ regions in our sample are necessary to disentangle
the effects of feedback mechanisms such as outflows from properties
like rotation.

\section{Summary}\label{sec:summary}

Over the last decade we have doubled the number of known \hii\ regions
in the Milky Way by detecting RRL emission at 4--10\ghz\ toward
candidate \hii\ regions that were identified by their mid-infrared and
radio continuum morphology.  This was achieved by the improved
spectral sensitivity of the GBT in the northern sky (HRDS) and the
ATCA in the southern sky (SHRDS).  A total of $\sim 1400$ \hii\
regions have been discovered and the current census now includes all
nebulae ionized by a single O9$\,$V-type star at a distance of
$\sim 10$\kpc.  We also used the VLA to observe a subset of the HRDS
to derive accurate electron temperatures which are a proxy for
metallicity.

Here we use ATCA and VLA RRL data to create a sample of 425
independent observations of 374 \hii\ regions that are suitable for
kinematic analysis.  We select sources that are relatively well
isolated and have a single RRL profile with a SNR greater than 10.  We
perform Gaussian fits of the RRL position-position-velocity data cubes
and discover velocity gradients in about 40\% of the nebulae.
Velocity gradient magnitudes range between about 5 and 200\msarcsec.
We also detect centrally peaked RRL widths in about 15\% of the
sources.  The velocity gradient position angles appear to be random on
the sky with no favored orientation with respect to the Galactic
Plane.

To better interpret these results we develop simulations to produce
synthetic RRL observations toward model nebulae.  Assuming a
homogeneous, isothermal spherical nebula we perform the radiative
transfer of RRL and continuum emission on the sky and convolve the
resulting brightness distribution with a telescope antenna pattern to
produce synthetic RRL data cubes.  We consider two dynamical models:
bipolar outflow and solid body rotation.  Both dynamical models
produce velocity gradients that have similar characteristics for
unresolved nebulae.  The solid body rotation model produces both the
full range of velocity gradient magnitudes and centrally peaked line
widths.  In contrast, the bipolar outflow model cannot produce the
observed velocity gradients $> 40$\msarcsec\ nor the centrally peaked
line widths for the simulations discussed here.  We therefore favor
the solid body rotation model, but we cannot rule out that bipolar
outflows are causing the velocity gradients in some of our sources.
There are several examples in our sample and in the literature of
\hii\ regions that show evidence for both rotation and outflow.
Observations with higher spatial resolution are required to
distinguish between these models for many of our sources.

\acknowledgments

The Australia Telescope Compact Array is part of the Australia
Telescope National Facility, which is funded by the Australian
Government for operation as a National Facility managed by CSIRO.  The
Green Bank Observatory and National Radio Astronomy Observatory are
facilities of the National Science Foundation operated under
cooperative agreement by Associated Universities, Inc.  This work is
supported by NSF grants AST1812639 to LDA and AST1714688 to TMB.  This
research has made use of NASA’s Astrophysics Data System Bibliographic
Services.

\vspace{5mm}
\facilities{ATCA, VLA}

\software{Astropy (Astropy Collaboration et a. 2013), Matplolib
  \citep{hunter07}, NumPY \& SciPy \citep{vanderwalt11}}

\appendix

\section{Fitting Models to Correlated Image Data}\label{sec:mle}

A model \(M\) predicts
\(\vect{y} = (y_0, y_1, \dots, y_N)\) given independent data
\(\vect{x} = (x_0, x_y, \dots, x_N)\) and model parameters
\(\vect{\theta}\). The generalized Gaussian log-likelihood function is
\begin{equation}
  \ln L = -0.5N\ln 2\pi - 0.5\ln|\vect{\Sigma}| - 0.5\left(\vect{y} - M(\vect{x}, \vect{\theta})\right)^T \vect{\Sigma}^{-1} \left(\vect{y} - M(\vect{x}, \vect{\theta})\right),  \label{eq:likelihood}
\end{equation}
where \(\vect{\Sigma}\) is the covariance matrix.

The generalized Gaussian likelihood function is simplified under the
assumption of independent and identically distributed (i.d.d.)
observed data.  In this scenario, the covariance matrix has zeros
every except along the diagonal, where \(\Sigma_{i,i} = \sigma^2_{y,
  i}\) and \(\vect{\sigma^2_y} = (\sigma^2_{y,0}, \sigma^2_{y, 1},
\dots, \sigma^2_{y, N})\) are the observed data variances. The
determinant of this covariance matrix is \(|\vect{\Sigma}| =
\sum^N\sigma^2_{y, i}\) and so the log-likelihood function is
\begin{equation}
  \ln L = -0.5N\ln 2\pi - 0.5\sum^N\sigma^2_{y, i} - 0.5\sum^N\frac{\left(y_i - M(x_i, \vect{\theta})\right)^2}{\sigma^2_{y, i}}.
\end{equation}
Maximizing this log-likelihood function is equivalent to reducing the last
term, which is the sum of the squared residuals weighted by the
observed data variances.

In general, the covariance matrix is populated with elements
\(\Sigma_{i,j} = \rho_{i,j}\sigma_i\sigma_j\) where \(\sigma_i\) and
\(\sigma_j\) are the standard deviations of the observed data \(y_i\)
and \(y_j\), respectively, and \(\rho_{i,j}\) is the correlation
coefficient between \(y_i\) and \(y_j\). In an image with a Gaussian
resolution element, the correlation coefficient between pixel \(i\)
and pixel \(j\) is
\begin{equation}
  \rho_{i,j} = \exp\left[-A\Delta x^2 - 2B\Delta x\Delta y - C\Delta y^2\right]\,,
\end{equation}
where
\begin{align}
  A & = \frac{\cos^2\phi}{2\theta_{\rm maj}^2} + \frac{\sin^2\phi}{2\theta_{\rm min}^2}\,, \nonumber \\
  B & = -\frac{\sin(2\phi)}{4\theta_{\rm maj}^2} + \frac{\sin(2\phi)}{4\theta_{\rm min}^2}\,, \nonumber \\
  C & = \frac{\sin^2\phi}{2\theta_{\rm maj}^2} + \frac{\cos^2\phi}{2\theta_{\rm min}^2} \,, \nonumber
\end{align}
\(\Delta x\) and \(\Delta y\) are the separations between the \(i\)th
and \(j\)th pixels in the east-west and north-south directions,
respectively, and \(\theta_{\rm maj}\), \(\theta_{\rm min}\), and
\(\phi\) are the synthesized beam major axis, minor axis, and
north-through-east position angle, respectively \citep[see][]{wenger19a}.

The generalized covariance matrix is typically ill-conditioned; the
determinant and inverted matrix are susceptible to numerical precision
loss. Therefore, we truncate the correlation coefficient at \(\rho =
0.5\), which is equivalent to the expected correlation between two
points separated by half of the half-power beam-width.

For a polynomial model, \(M(\vect{x}, \vect{\theta}) =
\vect{D}\vect{\theta}\) where \(\vect{D}\) is the design matrix with
elements \(D_{i,j} = x_i^j\). With this notation, the log-likelihood
function is
\begin{equation}
  \ln L = -0.5(\vect{y} - \vect{D}\vect{\theta})^T\vect{\Sigma}^{-1}(\vect{y} - \vect{D}\vect{\theta}) + const.
\end{equation}
Maximizing this function is identical to generalized least
squares. Equating the derivative of the log-likelihood with respect to
\(\vect{\theta}\) to zero, we solve for the parameters
\(\hat{\vect{\theta}}\) that maximize the log-likelihood:
\begin{align}
  \frac{\partial\,({\rm ln}\,L)}{\partial \vect{\theta}} = 0 & = -\vect{D}^T\vect{\Sigma}^{-1}(\vect{y} - \vect{D}\hat{\vect{\theta}}) \nonumber \\
  & = -\vect{D}^T\vect{\Sigma}^{-1}\vect{y} + \vect{D}^T\vect{\Sigma}^{-1}\vect{D}\hat{\vect{\theta}} \nonumber \\
  \hat{\vect{\theta}} & = (\vect{D}^T\vect{\Sigma}^{-1}\vect{D})^{-1}\vect{D}^T\vect{\Sigma}^{-1}\vect{y}.
\end{align}

\subsection{Velocity Gradient Fit to a Plane Model}\label{sec:plane}

To characterize the observed velocity gradients we fit the
$V_{\rm LSR}$ image to a plane defined by
\begin{equation}
z = c_{0} + c_{1}\,x + c_{2}\,y,
\end{equation}
where ($x,y$) are the pixel coordinates on the sky, $z$ is the LSR
velocity, and ($c_{0},c_{1},c_{2}$) are constants.  The magnitude of
the velocity gradient is
\begin{equation}
\nabla V_{\rm LSR} = \sqrt{c_{1}^2 + c_{2}^2},
\end{equation}
and the position angle,
\begin{equation}
{\rm PA} = {\rm arctan}(c_{2}/c_{1}).  
\end{equation}
The offset velocity is given by $c_{0}$.  We use Maximum Likelihood
Estimation as described in \S{\ref{sec:mle}} to determine the best fit
taking into account the errors in $V_{\rm LSR}$ and that nearby pixels
are correlated; that is, there are several pixels across the
synthesized HPBW.  The position angle is defined to be zero along the
north direction, increasing eastward in Equatorial coordinates.

\section{\hii\ Region Model}\label{sec:model}

To explore the effects of different dynamical models on the observed
RRL kinematics, we develop a numerical code in Python to simulation a
RRL and continuum emission observation (Ref, TBD).  We assume a
homogeneous, isothermal, spherical nebula consisting of fully ionized
hydrogen with electron density $n_{\rm e}$ and electron temperature
$T_{\rm e}$.  In practice the electron density and temperature vary
within an \hii\ region \citep{roelfsema92, balser18}, but here we keep
the models simple.  We include RRL and free-free continuum emission
assuming local thermodynamic equilibrium (LTE).  Following
\citet{condon16}, the RRL absorption coefficient as a function of
frequency, $\nu$, is given by
\begin{equation}
\kappa_{\rm L}(\nu) = \frac{c^2}{8\pi \nu_{\rm o}^2}\frac{g_{\rm u}}{g_{\rm l}}n_{\rm l}A_{\rm ul}\left[ 1 - {\rm exp}\left(-\frac{h\nu_{\rm o}}{kT_{\rm e}}\right) \right] \phi(\nu),
\end{equation}
where $c$ is the speed of light, $h$ is Planck's constant, $k$ is
Boltzmann's constant, $\nu_{\rm o}$ is the frequency of the RRL
transition, $A_{\rm ul}$ is the spontaneous emission rate from the
upper to the lower state, and $n_{\rm l}$ is the number density in the
lower state.  The spontaneous emission rate can be approximated as
\begin{equation}
A_{\rm n+1,n} = \left(\frac{64\pi^{6} m_{\rm e} e^{10}}{3 c^{3} h^{6}}\right) \frac{1}{n^{5}},
\end{equation}
where $m_{\rm e}$ is the electron mass and n is the principal quantum
number \citep{condon16}.  The statistical weights for hydrogen are
given by $g_{\rm n} = 2{\rm n}^{2}$.  We assume a Gaussian line
profile, $\phi(\nu)$, that is broadened by thermal motions of the gas
characterized by the FWHM line width, $\Delta V_{\rm t}$.  A
non-thermal component, $\Delta V_{\rm nt}$, is included to account for
any turbulent motions and is added in quadrature to the thermal width
to derive the total line width.

\hii\ regions also produce continuous, free-free thermal
bremsstrahlung emission.  From \citet{condon16}, the free-free
absorption coefficient is given by
\begin{equation}
\kappa_{\rm C}(\nu) = \frac{1}{\nu^{2}T_{\rm e}^{3/2}}\left[ \frac{Z^{2}e^{6}}{c}n_{\rm e}n_{\rm i}\frac{1}{\sqrt{2\pi(m_{\rm e}k)^{3}}}\right] \frac{\pi^{2}}{4}ln\left( \frac{b_{\rm max}}{b_{\rm min}}\right) ,
\end{equation}
where $Z$ is the effective nuclear charge number, $e$ is the charge,
$n_{\rm i}$ is the ion number density, and $b_{\rm min}$ and
$b_{\rm min}$ are the minimum and maximum impact parameters,
respectively.  Following \citet{condon16}, we estimate the impact
parameter ratio by
\begin{equation}
\frac{b_{\rm max}}{b_{\rm min}} = \left(\frac{3 k T_{\rm e}}{m_{\rm e}}\right)^{3/2}\frac{m_{\rm e}}{2 \pi Z e^{2} \nu} .
\end{equation}

Assuming the Rayleigh-Jeans limit, $h\nu \ll kT_{\rm e}$, the
brightness temperature is given by
\begin{equation}
T_{\rm B} = T_{\rm e}(1 - {\rm e}^{-\tau}),
\end{equation}
where $\tau$ is the optical depth, $\tau = \int \kappa\,d\ell$, and
$d\ell$ is the path length through the nebula.  We perform the
radiative transfer through the nebula to determine the brightness
distribution on the sky (\jya) as a function of frequency.  The
spectral noise is modeled as random (Gaussian) noise with a specified
RMS.  Synthetic spectra are produced by convolving the brightness
distribution on the sky with a telescope beam to determine the
observed brightness distribution (\jyb).  We assume a Gaussian
telescope beam shape with a specified HPBW.  The nebular angular size
is given by $\theta_{\rm s} = D/R_{\rm Sun}$, where $D$ is the
diameter of the spherical nebula and $R_{\rm Sun}$ is the distance.

The model \hii\ region properties are listed in Table~\ref{tab:hii}
with their default values.  Unless noted otherwise we use the default
values for all simulations.  In addition to the thermal and
non-thermal motions that are constant within the nebula, we include
bulk motions that are described by a dynamical model.  Below we
consider bipolar outflow and solid body rotation models.

\begin{deluxetable}{lccl}
\tablecaption{Model \hii\ Region Properties \label{tab:hii}}
\tablehead{
\colhead{Property} & \colhead{Symbol} & \colhead{Default Value} &
\colhead{Comment} 
}
\startdata
Electron density & $n_{\rm e}$ & 250\percc\ & Homogeneous nebula.\\
Electron temperature & $T_{\rm e}$ & 8000\kel\ & Isothermal nebula. \\
Non-thermal line width & $\Delta V_{\rm nt}$ & 15\kms\ & \\
Diameter & $D$ & 2\pc\ & Spherical nebula. \\
Distance & $R_{\rm Sun}$ & 5\kpc\ & \\
RRL & Hn$\alpha$ & H85$\alpha$ & $\Delta{\rm n} = 1$ correspond to $\alpha$ transitions. \\
RMS spectral noise & $N$ & 0.001\mjya\ & Gaussian noise. \\
Telescope HPBW & $\theta_{\rm b}$ & 90\arcsec\ & Gaussian profile. \\
Outflow speed & $V_{\rm o}$ & \dots & Bipolar outflow model. \\
Outflow opening angle & $\Theta_{\rm o}$ & \dots & Bipolar outflow model. \\
Rotation equatorial speed & $V_{\rm eq}$ & \dots & Solid body rotation model; speed at the equator. \\
Line-of-sight angle & $\Phi_{\rm los}$ & \dots & Bipolar outflow and solid body rotation models. \\
Sky position angle & $\Phi_{\rm sky}$ & \dots & Bipolar outflow and solid body rotation models. \\
\enddata
%\tablecomments{} 
%\tablenotetext{a}{}
\end{deluxetable}

\subsection{Bipolar Outflow}\label{sec:bipolar}

Outflows are a common phenomena in astrophysics and may explain the
observed velocity gradients.  We assume that the bipolar outflow
emanates radially from the center of the \hii\ region sphere and is
defined by a constant outflow speed, $V_{\rm o}$, with opening angle,
$\Theta_{\rm o}$ (see Figure~\ref{fig:cartoon}).  The outflow
orientation is defined by two angles: the position angle relative to
the line-of-sight, $\Phi_{\rm los}$, and the position angle on the
sky, $\Phi_{\rm sky}$. The angle $\Phi_{\rm los}$ is zero normal to
the observer and 90\degree\ along the line-of-sight.  The angle
$\Phi_{\rm sky}$ is defined to be zero when oriented north and
increases to the east.

\subsection{Solid Body Rotation}\label{sec:rotation}

\begin{figure}
  \centering
  \includegraphics[angle=0,scale=0.6]{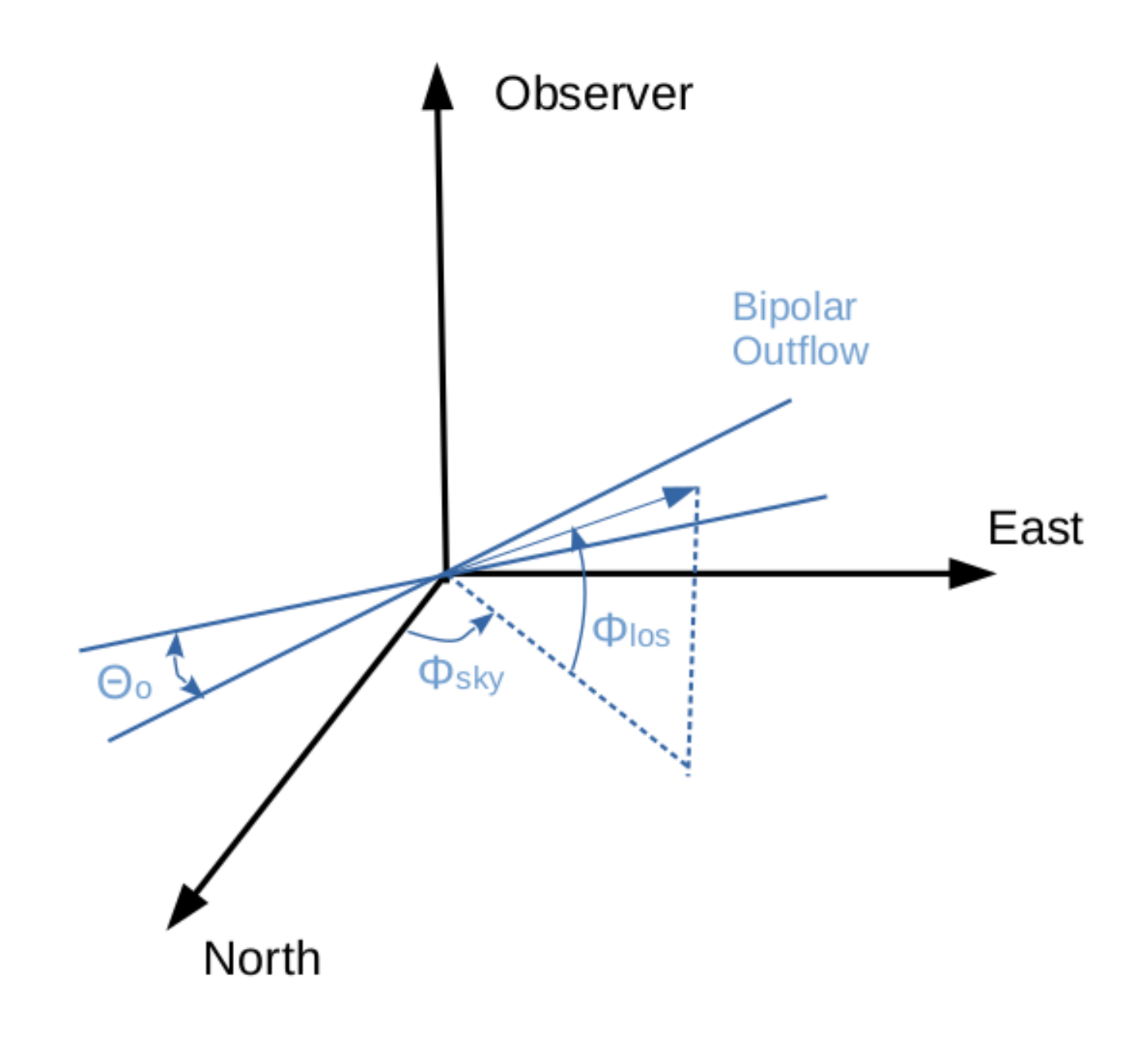}
  \includegraphics[angle=0,scale=0.6]{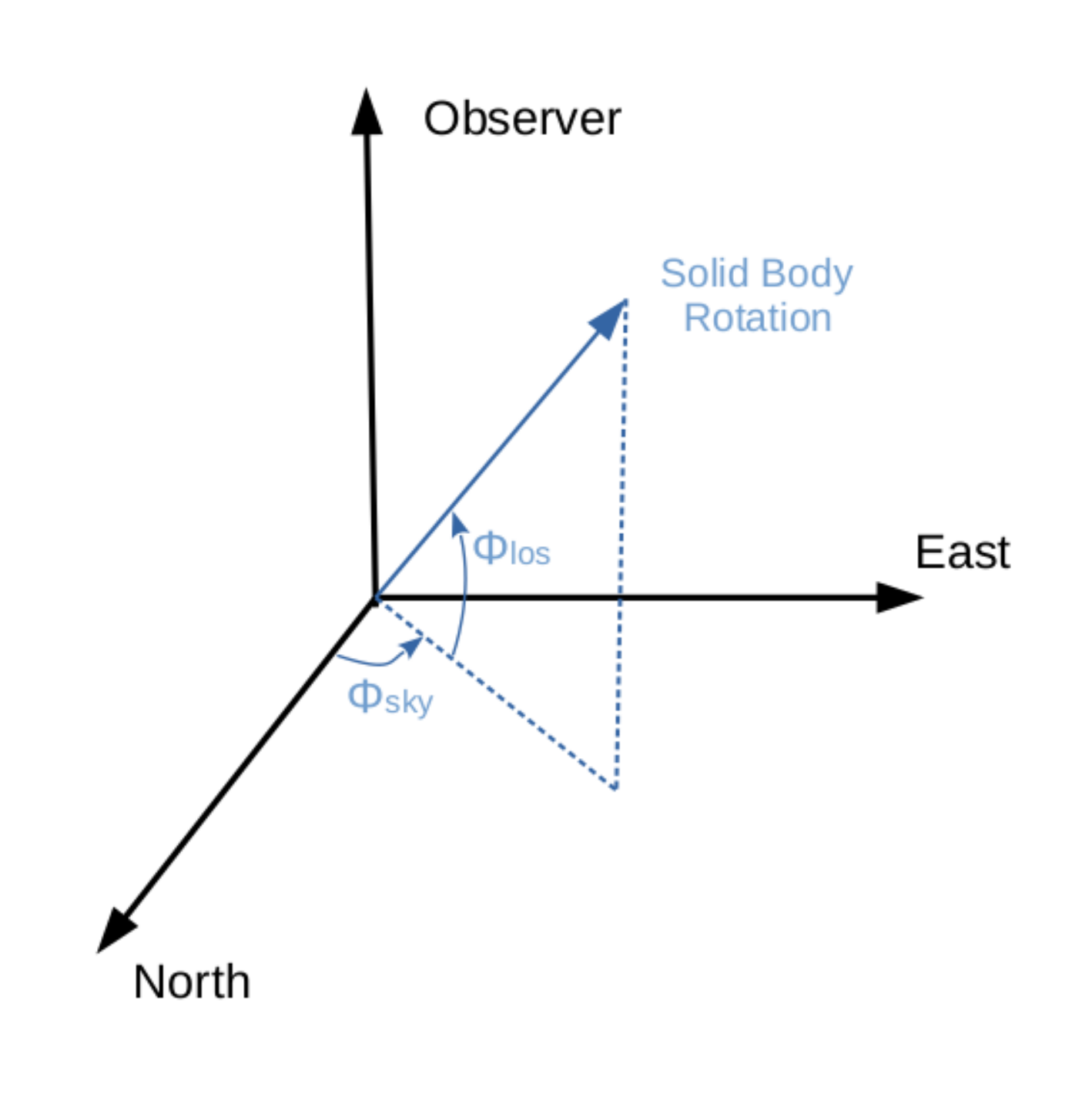}
  \caption{Geometry of the bipolar outflow (left) and solid body
    rotation (right) dynamical models.  The observer is located along
    the z-axis, looking down on the x-y plane.  The angle of the
    outflow or rotation axis relative to the line-of-sight,
    $\Phi_{\rm los}$, is 90\degree\ when pointing towards the
    observer.  The position angle of the outflow or rotation axis on
    the sky, $\Phi_{\rm sky}$, is zero when pointing to the north.
    For the bipolar outflow model the opening angle of the outflow,
    assumed to be symmetric, is given by $\Theta_{\rm o}$.}
  \label{fig:cartoon}
\end{figure}

Rotation is a common property in astrophysics and we might expect that
all \hii\ region complexes inherit some angular momentum, and thus
rotation, from their parent molecular cloud.  Here we explore solid
body rotation for simplicity, but differential rotation is also
possible.  The magnitude of the solid body rotation is defined by the
equatorial speed, $V_{\rm eq}$, and the orientation of the angular
momentum vector is defined by the two position angles:
$\Phi_{\rm los}$ and $\Phi_{\rm sky}$ (see Figure~\ref{fig:cartoon}).
The rotation axis is the direction of the angular momentum vector, so
the angle $\Phi_{\rm los}$ is zero normal to the observer and
90\degree\ along the line-of-sight.  The angle $\Phi_{\rm sky}$ is
defined to be zero when oriented north and increases to the east.

\bibliography{ref}

\newcommand{\noop}[1]{}
\begin{thebibliography}{}
\expandafter\ifx\csname natexlab\endcsname\relax\def\natexlab#1{#1}\fi
\providecommand{\url}[1]{\href{#1}{#1}}
\providecommand{\dodoi}[1]{doi:~\href{http://doi.org/#1}{\nolinkurl{#1}}}
\providecommand{\doeprint}[1]{\href{http://ascl.net/#1}{\nolinkurl{http://ascl.net/#1}}}
\providecommand{\doarXiv}[1]{\href{https://arxiv.org/abs/#1}{\nolinkurl{https://arxiv.org/abs/#1}}}

\bibitem[{{Ali} {et~al.}(2018){Ali}, {Harries}, \& {Douglas}}]{ali18}
{Ali}, A., {Harries}, T.~J., \& {Douglas}, T.~A. 2018, \mnras, 477, 5422,
  \dodoi{10.1093/mnras/sty1001}

\bibitem[{{Anderson} {et~al.}(2015){Anderson}, {Armentrout}, {Johnstone},
  {Bania}, {Balser}, {Wenger}, \& {Cunningham}}]{anderson15}
{Anderson}, L.~D., {Armentrout}, W.~P., {Johnstone}, B.~M., {et~al.} 2015,
  \apjs, 221, 26, \dodoi{10.1088/0067-0049/221/2/26}

\bibitem[{{Anderson} {et~al.}(2018){Anderson}, {Armentrout}, {Luisi}, {Bania},
  {Balser}, \& {Wenger}}]{anderson18}
{Anderson}, L.~D., {Armentrout}, W.~P., {Luisi}, M., {et~al.} 2018, \apjs, 234,
  33, \dodoi{10.3847/1538-4365/aa956a}

\bibitem[{{Anderson} {et~al.}(2011){Anderson}, {Bania}, {Balser}, \&
  {Rood}}]{anderson11}
{Anderson}, L.~D., {Bania}, T.~M., {Balser}, D.~S., \& {Rood}, R.~T. 2011,
  \apjs, 194, 32, \dodoi{10.1088/0067-0049/194/2/32}

\bibitem[{{Andr{\'e}} {et~al.}(2014){Andr{\'e}}, {Di Francesco},
  {Ward-Thompson}, {Inutsuka}, {Pudritz}, \& {Pineda}}]{andre14}
{Andr{\'e}}, P., {Di Francesco}, J., {Ward-Thompson}, D., {et~al.} 2014, in
  Protostars and Planets VI, ed. H.~{Beuther}, R.~S. {Klessen}, C.~P.
  {Dullemond}, \& T.~{Henning}, 27,
  \dodoi{10.2458/azu\_uapress\_9780816531240-ch002}

\bibitem[{{Andr{\'e}} {et~al.}(2010){Andr{\'e}}, {Men'shchikov}, {Bontemps},
  {K{\"o}nyves}, {Motte}, {Schneider}, {Didelon}, {Minier}, {Saraceno},
  {Ward-Thompson}, {di Francesco}, {White}, {Molinari}, {Testi}, {Abergel},
  {Griffin}, {Henning}, {Royer}, {Mer{\'\i}n}, {Vavrek}, {Attard},
  {Arzoumanian}, {Wilson}, {Ade}, {Aussel}, {Baluteau}, {Benedettini},
  {Bernard}, {Blommaert}, {Cambr{\'e}sy}, {Cox}, {di Giorgio}, {Hargrave},
  {Hennemann}, {Huang}, {Kirk}, {Krause}, {Launhardt}, {Leeks}, {Le Pennec},
  {Li}, {Martin}, {Maury}, {Olofsson}, {Omont}, {Peretto}, {Pezzuto}, {Prusti},
  {Roussel}, {Russeil}, {Sauvage}, {Sibthorpe}, {Sicilia-Aguilar}, {Spinoglio},
  {Waelkens}, {Woodcraft}, \& {Zavagno}}]{andre10}
{Andr{\'e}}, P., {Men'shchikov}, A., {Bontemps}, S., {et~al.} 2010, \aap, 518,
  L102, \dodoi{10.1051/0004-6361/201014666}

\bibitem[{{Arquilla} \& {Goldsmith}(1985)}]{arquilla85}
{Arquilla}, R., \& {Goldsmith}, P.~F. 1985, \apj, 297, 436,
  \dodoi{10.1086/163542}

\bibitem[{{Balser}(2006)}]{balser06}
{Balser}, D.~S. 2006, \aj, 132, 2326, \dodoi{10.1086/508515}

\bibitem[{{Balser} \& {Bania}(2018)}]{balser18}
{Balser}, D.~S., \& {Bania}, T.~M. 2018, \aj, 156, 280,
  \dodoi{10.3847/1538-3881/aaeb2b}

\bibitem[{{Balser} {et~al.}(1997){Balser}, {Bania}, {Rood}, \&
  {Wilson}}]{balser97}
{Balser}, D.~S., {Bania}, T.~M., {Rood}, R.~T., \& {Wilson}, T.~L. 1997, \apj,
  483, 320, \dodoi{10.1086/304248}

\bibitem[{{Balser} {et~al.}(2001){Balser}, {Goss}, \& {De Pree}}]{balser01}
{Balser}, D.~S., {Goss}, W.~M., \& {De Pree}, C.~G. 2001, \aj, 121, 371,
  \dodoi{10.1086/318028}

\bibitem[{{Bania} {et~al.}(2012){Bania}, {Anderson}, \& {Balser}}]{bania12}
{Bania}, T.~M., {Anderson}, L.~D., \& {Balser}, D.~S. 2012, \apj, 759, 96,
  \dodoi{10.1088/0004-637X/759/2/96}

\bibitem[{{Bania} {et~al.}(2010){Bania}, {Anderson}, {Balser}, \&
  {Rood}}]{bania10}
{Bania}, T.~M., {Anderson}, L.~D., {Balser}, D.~S., \& {Rood}, R.~T. 2010,
  \apjl, 718, L106, \dodoi{10.1088/2041-8205/718/2/L106}

\bibitem[{{Beckman} \& {Rela{\~n}o}(2004)}]{beckman04}
{Beckman}, J.~E., \& {Rela{\~n}o}, M. 2004, \apss, 292, 111,
  \dodoi{10.1023/B:ASTR.0000045006.88476.b6}

\bibitem[{{Belloche}(2013)}]{belloche13}
{Belloche}, A. 2013, in EAS Publications Series, Vol.~62, EAS Publications
  Series, ed. P.~{Hennebelle} \& C.~{Charbonnel}, 25--66,
  \dodoi{10.1051/eas/1362002}

\bibitem[{{Bodenheimer}(1995)}]{bodenheimer95}
{Bodenheimer}, P. 1995, \araa, 33, 199,
  \dodoi{10.1146/annurev.aa.33.090195.001215}

\bibitem[{{Bodenheimer} {et~al.}(1979){Bodenheimer}, {Tenorio-Tagle}, \&
  {Yorke}}]{bodenheimer79}
{Bodenheimer}, P., {Tenorio-Tagle}, G., \& {Yorke}, H.~W. 1979, \apj, 233, 85,
  \dodoi{10.1086/157368}

\bibitem[{{Bordalo} {et~al.}(2009){Bordalo}, {Plana}, \& {Telles}}]{bordalo09}
{Bordalo}, V., {Plana}, H., \& {Telles}, E. 2009, \apj, 696, 1668,
  \dodoi{10.1088/0004-637X/696/2/1668}

\bibitem[{{Bresolin} {et~al.}(2020){Bresolin}, {Rizzi}, {Ho}, {Terlevich},
  {Terlevich}, {Telles}, {Ch{\'a}vez}, {Basilakos}, \& {Plionis}}]{bresolin20}
{Bresolin}, F., {Rizzi}, L., {Ho}, I.~T., {et~al.} 2020, \mnras, 495, 4347,
  \dodoi{10.1093/mnras/staa1472}

\bibitem[{{Brown} {et~al.}(2017){Brown}, {Jordan}, {Dickey}, {Anderson},
  {Armentrout}, {Balser}, {Bania}, {Dawson}, {McClure-Griffiths}, \&
  {Wenger}}]{brown17}
{Brown}, C., {Jordan}, C., {Dickey}, J.~M., {et~al.} 2017, \aj, 154, 23,
  \dodoi{10.3847/1538-3881/aa71a7}

\bibitem[{{Caselli} {et~al.}(2002){Caselli}, {Benson}, {Myers}, \&
  {Tafalla}}]{caselli02}
{Caselli}, P., {Benson}, P.~J., {Myers}, P.~C., \& {Tafalla}, M. 2002, \apj,
  572, 238, \dodoi{10.1086/340195}

\bibitem[{{Chen} {et~al.}(2007){Chen}, {Launhardt}, \& {Henning}}]{chen07}
{Chen}, X., {Launhardt}, R., \& {Henning}, T. 2007, \apj, 669, 1058,
  \dodoi{10.1086/521868}

\bibitem[{{Churchwell}(2002)}]{churchwell02}
{Churchwell}, E. 2002, \araa, 40, 27,
  \dodoi{10.1146/annurev.astro.40.060401.093845}

\bibitem[{{Condon} \& {Ransom}(2016)}]{condon16}
{Condon}, J.~J., \& {Ransom}, S.~M. 2016, {Essential Radio Astronomy}

\bibitem[{{Dalgleish} {et~al.}(2018){Dalgleish}, {Longmore}, {Peters},
  {Henshaw}, {Veitch-Michaelis}, \& {Urquhart}}]{dalgleish18}
{Dalgleish}, H.~S., {Longmore}, S.~N., {Peters}, T., {et~al.} 2018, \mnras,
  478, 3530, \dodoi{10.1093/mnras/sty1109}

\bibitem[{{De Pree} {et~al.}(1994){De Pree}, {Goss}, {Palmer}, \&
  {Rubin}}]{depree94}
{De Pree}, C.~G., {Goss}, W.~M., {Palmer}, P., \& {Rubin}, R.~H. 1994, \apj,
  428, 670, \dodoi{10.1086/174277}

\bibitem[{{De Pree} {et~al.}(1995){De Pree}, {Rodriguez}, {Dickel}, \&
  {Goss}}]{depree95}
{De Pree}, C.~G., {Rodriguez}, L.~F., {Dickel}, H.~R., \& {Goss}, W.~M. 1995,
  \apj, 447, 220, \dodoi{10.1086/175868}

\bibitem[{{Dyson} \& {Williams}(1997)}]{dyson97}
{Dyson}, J.~E., \& {Williams}, D.~A. 1997, {The physics of the interstellar
  medium}, \dodoi{10.1201/9780585368115}

\bibitem[{{Ferland}(2001)}]{ferland01}
{Ferland}, G.~J. 2001, \pasp, 113, 41, \dodoi{10.1086/317983}

\bibitem[{{Fleck} \& {Clark}(1981)}]{fleck81}
{Fleck}, R.~C., J., \& {Clark}, F.~O. 1981, \apj, 245, 898,
  \dodoi{10.1086/158866}

\bibitem[{{Garay} {et~al.}(1986){Garay}, {Rodriguez}, \& {van
  Gorkom}}]{garay86}
{Garay}, G., {Rodriguez}, L.~F., \& {van Gorkom}, J.~H. 1986, \apj, 309, 553,
  \dodoi{10.1086/164624}

\bibitem[{{Goodman} {et~al.}(1993){Goodman}, {Benson}, {Fuller}, \&
  {Myers}}]{goodman93}
{Goodman}, A.~A., {Benson}, P.~J., {Fuller}, G.~A., \& {Myers}, P.~C. 1993,
  \apj, 406, 528, \dodoi{10.1086/172465}

\bibitem[{{Hsieh} {et~al.}(2021){Hsieh}, {Arce}, {Mardones}, {Kong}, \&
  {Plunkett}}]{hsieh21}
{Hsieh}, C.-H., {Arce}, H.~G., {Mardones}, D., {Kong}, S., \& {Plunkett}, A.
  2021, \apj, 908, 92, \dodoi{10.3847/1538-4357/abd034}

\bibitem[{{Hunter}(2007)}]{hunter07}
{Hunter}, J.~D. 2007, Computing in Science Engineering, 9, 90,
  \dodoi{10.1109/MCSE.2007.55}

\bibitem[{{Imara} \& {Blitz}(2011)}]{imara11}
{Imara}, N., \& {Blitz}, L. 2011, \apj, 732, 78,
  \dodoi{10.1088/0004-637X/732/2/78}

\bibitem[{{Jaffe} {et~al.}(2005){Jaffe}, {Zhu}, {Lacy}, {Richter}, \&
  {Greathouse}}]{jaffe05}
{Jaffe}, D., {Zhu}, Q., {Lacy}, J., {Richter}, M., \& {Greathouse}, T. 2005, in
  High Resolution Infrared Spectroscopy in Astronomy, 162--167,
  \dodoi{10.1007/10995082_26}

\bibitem[{{Jeffreson} {et~al.}(2020){Jeffreson}, {Kruijssen}, {Keller},
  {Chevance}, \& {Glover}}]{jeffreson20}
{Jeffreson}, S. M.~R., {Kruijssen}, J.~M.~D., {Keller}, B.~W., {Chevance}, M.,
  \& {Glover}, S. C.~O. 2020, \mnras, \dodoi{10.1093/mnras/staa2127}

\bibitem[{Kass \& Raftery(1995)}]{kass95}
Kass, R.~E., \& Raftery, A.~E. 1995, Journal of the American Statistical
  Association, 90, 773, \dodoi{10.1080/01621459.1995.10476572}

\bibitem[{{Keto} \& {Wood}(2006)}]{keto06}
{Keto}, E., \& {Wood}, K. 2006, \apj, 637, 850, \dodoi{10.1086/498611}

\bibitem[{{Koch} \& {Rosolowsky}(2015)}]{koch15}
{Koch}, E.~W., \& {Rosolowsky}, E.~W. 2015, \mnras, 452, 3435,
  \dodoi{10.1093/mnras/stv1521}

\bibitem[{{Lebouteiller} {et~al.}(2012){Lebouteiller}, {Cormier}, {Madden},
  {Galliano}, {Indebetouw}, {Abel}, {Sauvage}, {Hony}, {Contursi}, {Poglitsch},
  {R{\'e}my}, {Sturm}, \& {Wu}}]{lebouteiller12}
{Lebouteiller}, V., {Cormier}, D., {Madden}, S.~C., {et~al.} 2012, \aap, 548,
  A91, \dodoi{10.1051/0004-6361/201218859}

\bibitem[{{Lopez} {et~al.}(2014){Lopez}, {Krumholz}, {Bolatto}, {Prochaska},
  {Ramirez-Ruiz}, \& {Castro}}]{lopez14}
{Lopez}, L.~A., {Krumholz}, M.~R., {Bolatto}, A.~D., {et~al.} 2014, \apj, 795,
  121, \dodoi{10.1088/0004-637X/795/2/121}

\bibitem[{{Luisi} {et~al.}(2021){Luisi}, {Anderson}, {Schneider}, {Simon},
  {Kabanovic}, {G{\"u}sten}, {Zavagno}, {Broos}, {Buchbender}, {Guevara},
  {Jacobs}, {Justen}, {Klein}, {Linville}, {R{\"o}llig}, {Russeil}, {Stutzki},
  {Tiwari}, {Townsley}, \& {Tielens}}]{luisi21}
{Luisi}, M., {Anderson}, L.~D., {Schneider}, N., {et~al.} 2021, arXiv e-prints,
  arXiv:2104.04568.
\newblock \doarXiv{2104.04568}

\bibitem[{{Markwardt}(2009)}]{markwardt09}
{Markwardt}, C.~B. 2009, in Astronomical Society of the Pacific Conference
  Series, Vol. 411, Astronomical Data Analysis Software and Systems XVIII, ed.
  D.~A. {Bohlender}, D.~{Durand}, \& P.~{Dowler}, 251.
\newblock \doarXiv{0902.2850}

\bibitem[{{Mathieu}(2004)}]{mathieu04}
{Mathieu}, R.~D. 2004, in Stellar Rotation, ed. A.~{Maeder} \& P.~{Eenens},
  Vol. 215, 113

\bibitem[{{McLeod} {et~al.}(2019){McLeod}, {Dale}, {Evans}, {Ginsburg},
  {Kruijssen}, {Pellegrini}, {Ramsay}, \& {Testi}}]{mcleod19}
{McLeod}, A.~F., {Dale}, J.~E., {Evans}, C.~J., {et~al.} 2019, \mnras, 486,
  5263, \dodoi{10.1093/mnras/sty2696}

\bibitem[{{Mermilliod} \& {Garc{\'\i}a}(2001)}]{mermilliod01}
{Mermilliod}, J.-C., \& {Garc{\'\i}a}, B. 2001, in The Formation of Binary
  Stars, ed. H.~{Zinnecker} \& R.~{Mathieu}, Vol. 200, 191

\bibitem[{{Misugi} {et~al.}(2019){Misugi}, {Inutsuka}, \&
  {Arzoumanian}}]{misugi19}
{Misugi}, Y., {Inutsuka}, S.-i., \& {Arzoumanian}, D. 2019, \apj, 881, 11,
  \dodoi{10.3847/1538-4357/ab2382}

\bibitem[{{Morris} {et~al.}(1974){Morris}, {Zuckerman}, {Turner}, \&
  {Palmer}}]{morris74}
{Morris}, M., {Zuckerman}, B., {Turner}, B.~E., \& {Palmer}, P. 1974, \apjl,
  192, L27, \dodoi{10.1086/181581}

\bibitem[{{Mouschovias}(1975)}]{mouschovias75}
{Mouschovias}, T.~C. 1975, PhD thesis, California Univ., Berkeley.

\bibitem[{{Mufson} \& {Liszt}(1977)}]{mufson77}
{Mufson}, S.~L., \& {Liszt}, H.~S. 1977, \apj, 212, 664, \dodoi{10.1086/155089}

\bibitem[{{Nicholls} {et~al.}(2012){Nicholls}, {Dopita}, \&
  {Sutherland}}]{nicholls12}
{Nicholls}, D.~C., {Dopita}, M.~A., \& {Sutherland}, R.~S. 2012, \apj, 752,
  148, \dodoi{10.1088/0004-637X/752/2/148}

\bibitem[{{Oey}(1996)}]{oey96}
{Oey}, M.~S. 1996, \apj, 465, 231, \dodoi{10.1086/177415}

\bibitem[{{Okamoto} {et~al.}(2003){Okamoto}, {Kataza}, {Yamashita}, {Miyata},
  {Sako}, {Takubo}, {Honda}, \& {Onaka}}]{okamoto03}
{Okamoto}, Y.~K., {Kataza}, H., {Yamashita}, T., {et~al.} 2003, \apj, 584, 368,
  \dodoi{10.1086/345540}

\bibitem[{{Peters} {et~al.}(2010{\natexlab{a}}){Peters}, {Banerjee}, {Klessen},
  {Mac Low}, {Galv{\'a}n-Madrid}, \& {Keto}}]{peters10a}
{Peters}, T., {Banerjee}, R., {Klessen}, R.~S., {et~al.} 2010{\natexlab{a}},
  \apj, 711, 1017, \dodoi{10.1088/0004-637X/711/2/1017}

\bibitem[{{Peters} {et~al.}(2010{\natexlab{b}}){Peters}, {Klessen}, {Mac Low},
  \& {Banerjee}}]{peters10c}
{Peters}, T., {Klessen}, R.~S., {Mac Low}, M.-M., \& {Banerjee}, R.
  2010{\natexlab{b}}, \apj, 725, 134, \dodoi{10.1088/0004-637X/725/1/134}

\bibitem[{{Peters} {et~al.}(2010{\natexlab{c}}){Peters}, {Mac Low}, {Banerjee},
  {Klessen}, \& {Dullemond}}]{peters10b}
{Peters}, T., {Mac Low}, M.-M., {Banerjee}, R., {Klessen}, R.~S., \&
  {Dullemond}, C.~P. 2010{\natexlab{c}}, \apj, 719, 831,
  \dodoi{10.1088/0004-637X/719/1/831}

\bibitem[{{Roelfsema} {et~al.}(1992){Roelfsema}, {Goss}, \&
  {Mallik}}]{roelfsema92}
{Roelfsema}, P.~R., {Goss}, W.~M., \& {Mallik}, D.~C.~V. 1992, \apj, 394, 188,
  \dodoi{10.1086/171570}

\bibitem[{{Rubin}(1985)}]{rubin85}
{Rubin}, R.~H. 1985, \apjs, 57, 349, \dodoi{10.1086/191007}

\bibitem[{{Rugel} {et~al.}(2019){Rugel}, {Rahner}, {Beuther}, {Pellegrini},
  {Wang}, {Soler}, {Ott}, {Brunthaler}, {Anderson}, {Mottram}, {Henning},
  {Goldsmith}, {Heyer}, {Klessen}, {Bihr}, {Menten}, {Smith}, {Urquhart},
  {Ragan}, {Glover}, {McClure-Griffiths}, {Bigiel}, \& {Roy}}]{rugel19}
{Rugel}, M.~R., {Rahner}, D., {Beuther}, H., {et~al.} 2019, \aap, 622, A48,
  \dodoi{10.1051/0004-6361/201834068}

\bibitem[{{Russeil} {et~al.}(2016){Russeil}, {Tig{\'e}}, {Adami}, {Anderson},
  {Schneider}, {Zavagno}, {Samal}, {Amram}, {Guennou}, {Le Coarer}, {Walsh},
  {Longmore}, \& {Purcell}}]{russeil16}
{Russeil}, D., {Tig{\'e}}, J., {Adami}, C., {et~al.} 2016, \aap, 587, A135,
  \dodoi{10.1051/0004-6361/201424484}

\bibitem[{{Spitzer}(1978)}]{spitzer78}
{Spitzer}, L. 1978, {Physical processes in the interstellar medium},
  \dodoi{10.1002/9783527617722}

\bibitem[{{Tobin} {et~al.}(2011){Tobin}, {Hartmann}, {Chiang}, {Looney},
  {Bergin}, {Chandler}, {Masqu{\'e}}, {Maret}, \& {Heitsch}}]{tobin11}
{Tobin}, J.~J., {Hartmann}, L., {Chiang}, H.-F., {et~al.} 2011, \apj, 740, 45,
  \dodoi{10.1088/0004-637X/740/1/45}

\bibitem[{{Torres-Flores} {et~al.}(2013){Torres-Flores}, {Barb{\'a}},
  {Ma{\'\i}z Apell{\'a}niz}, {Rubio}, {Bosch}, {H{\'e}nault-Brunet}, \&
  {Evans}}]{torres-flores13}
{Torres-Flores}, S., {Barb{\'a}}, R., {Ma{\'\i}z Apell{\'a}niz}, J., {et~al.}
  2013, \aap, 555, A60, \dodoi{10.1051/0004-6361/201220474}

\bibitem[{{van der Walt} {et~al.}(2011){van der Walt}, {Colbert}, \&
  {Varoquaux}}]{vanderwalt11}
{van der Walt}, S., {Colbert}, S.~C., \& {Varoquaux}, G. 2011, Computing in
  Science Engineering, 13, 22, \dodoi{10.1109/MCSE.2011.37}

\bibitem[{{Veilleux} {et~al.}(2005){Veilleux}, {Cecil}, \&
  {Bland-Hawthorn}}]{veilleux05}
{Veilleux}, S., {Cecil}, G., \& {Bland-Hawthorn}, J. 2005, \araa, 43, 769,
  \dodoi{10.1146/annurev.astro.43.072103.150610}

\bibitem[{{Welch} {et~al.}(1987){Welch}, {Dreher}, {Jackson}, {Terebey}, \&
  {Vogel}}]{welch87}
{Welch}, W.~J., {Dreher}, J.~W., {Jackson}, J.~M., {Terebey}, S., \& {Vogel},
  S.~N. 1987, Science, 238, 1550, \dodoi{10.1126/science.238.4833.1550}

\bibitem[{{Wenger} {et~al.}(2019{\natexlab{a}}){Wenger}, {Balser}, {Anderson},
  \& {Bania}}]{wenger19a}
{Wenger}, T.~V., {Balser}, D.~S., {Anderson}, L.~D., \& {Bania}, T.~M.
  2019{\natexlab{a}}, \apj, 887, 114, \dodoi{10.3847/1538-4357/ab53d3}

\bibitem[{{Wenger} {et~al.}(2019{\natexlab{b}}){Wenger}, {Dickey}, {Jordan},
  {Balser}, {Armentrout}, {Anderson}, {Bania}, {Dawson}, {McClure-Griffiths},
  \& {Shea}}]{wenger19b}
{Wenger}, T.~V., {Dickey}, J.~M., {Jordan}, C.~H., {et~al.} 2019{\natexlab{b}},
  \apjs, 240, 24, \dodoi{10.3847/1538-4365/aaf8ba}

\bibitem[{{Wenger} {et~al.}(2021){Wenger}, {Dawson}, {Dickey}, {Jordan},
  {McClure-Griffiths}, {Anderson}, {Armentrout}, {Balser}, \&
  {Bania}}]{wenger21}
{Wenger}, T.~V., {Dawson}, J.~R., {Dickey}, J.~M., {et~al.} 2021, arXiv
  e-prints, arXiv:2103.12199.
\newblock \doarXiv{2103.12199}

\bibitem[{{Zhu}(2006)}]{zhu06}
{Zhu}, Q. 2006, PhD thesis, The University of Texas at Austin

\end{thebibliography}

\end{document}